 \documentclass[final,3p,times,twocolumn]{elsarticle}

\newcommand{\MM}{\textbf M }

 \usepackage{graphicx}
 \usepackage{epstopdf}

\usepackage{amssymb}
\usepackage{amsmath}

\journal{Current Opinion in Solid State and Materials Science}

\begin{document}

\begin{frontmatter}



\title{The Magnetoelectric Effect in Transition Metal Oxides: \\Insights and the Rational Design of New Materials from First Principles}


\author[label1]{Turan Birol\fnref{label3}}
\author[label1,label2]{Nicole A.\ Benedek\fnref{label3}}
\author[label1] {Hena Das}
\author[label1] {Aleksander L. Wysocki}
\author[label1] {Andrew T.\ Mulder}
\author[label1] { Brian M.\ Abbett}
\author[label1] { Eva H.\ Smith}
\author[label1] { Saurabh Ghosh}
\author[label1] { Craig J.\ Fennie}
\address[label1]{School of Applied and Engineering Physics, Cornell University, Ithaca, New York 14853 USA}
\address[label2]{Materials Science and Engineering Program, The University of Texas at Austin, 1 University Station, Austin, Texas 78712 USA}
\fntext[label3]{These authors contributed equally}
\ead{fennie@cornell.edu}

\begin{abstract}
The search for materials displaying a large magnetoelectric effect has occupied researchers for many decades. The rewards could include not only advanced electronics technologies, but also fundamental insights concerning the dielectric and magnetic properties of condensed matter. In this article, we focus on the magnetoelectric effect in transition metal oxides and review the manner in which first-principles calculations have helped guide the search for (and increasingly, predicted) new materials and shed light on the microscopic mechanisms responsible for magnetoelectric phenomena.\\
\end{abstract}


\end{frontmatter}

\section{\label{intro} The challenge of discovering materials with large magnetoelectric responses -- First-principles theory to the rescue}

\subsection{Introduction}

\begin{figure}
  \begin{center}
    \includegraphics[width=0.8\hsize]{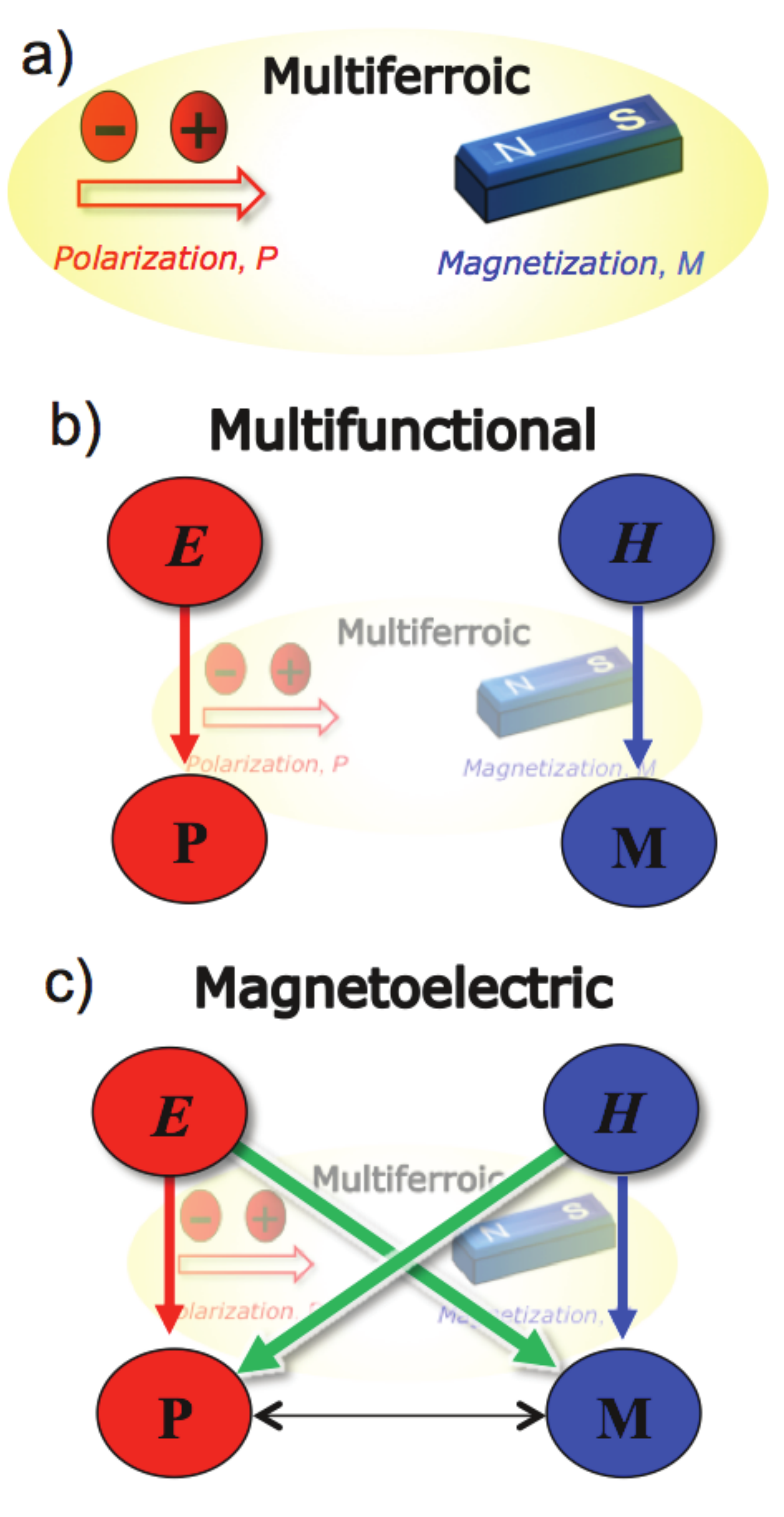}\\
        \includegraphics[width=0.9\hsize]{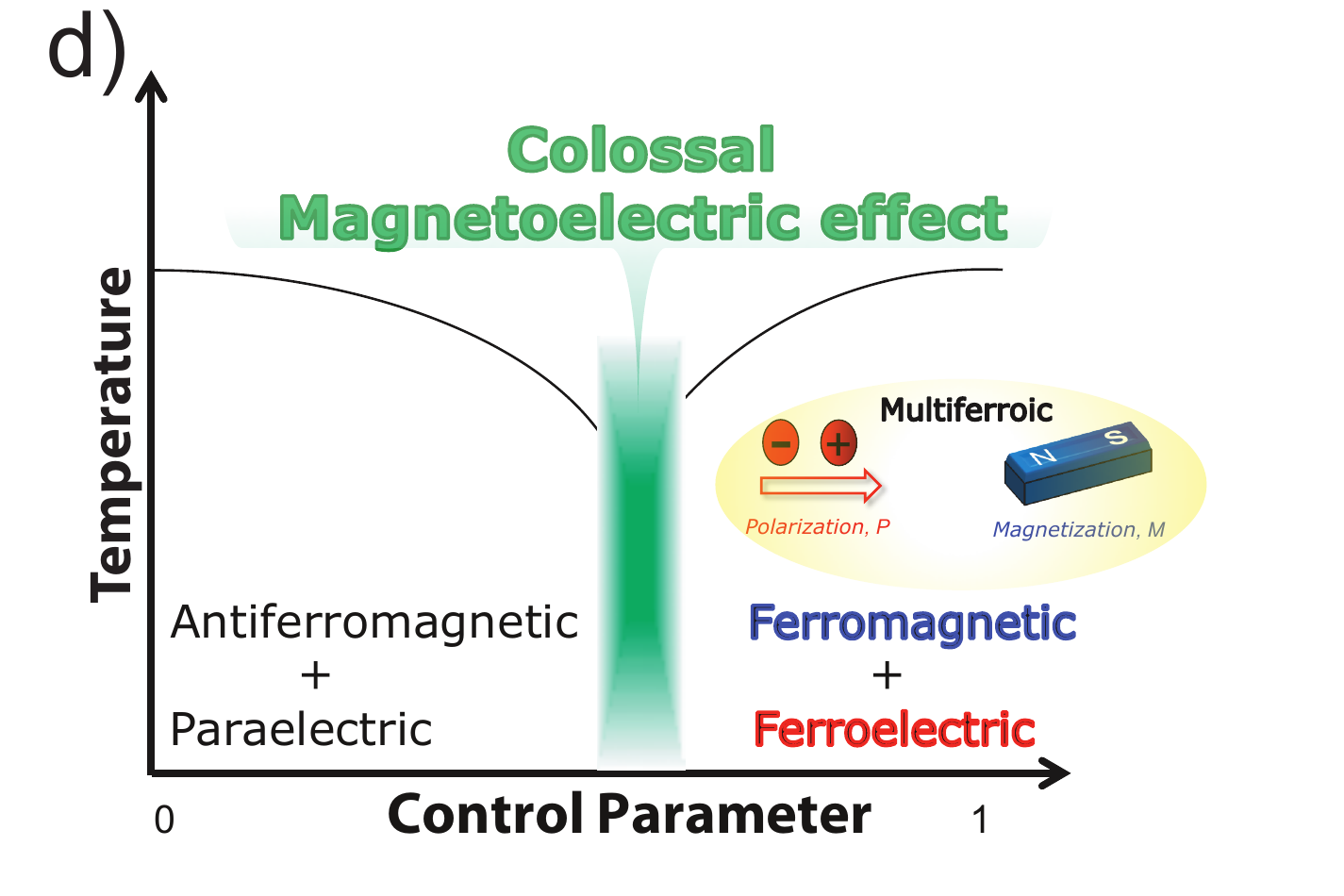}
  \end{center}
    \caption{(a) An example of a multiferroic are ferroelectric-ferromagnets, which have a spontaneous polarization and magnetization. (b) Multiferroics are multifunctional in that they respond in a useful way to more than one external perturbation, e..g, an electric field couples naturally to the electrical polarization, $P$, and a magnetic field couples naturally to the magnetization. (c) The generalized magnetoelectric effect: The Challenge is understanding and subsequently exploiting a microscopic mechanism that couples the electric and magnetic orderings at any order in the electric, $E$, and magnetic, $H$ fields. (d) Cartoon phase diagram representing the idea of phase competition. On the boundary between antiferromagnetic - paraelectric and ferromagnetic - ferroelectric phases, susceptibilities have divergent behavior and hence even small fields can induce colossal magnetoelectric effect.}
    \label{fig1}
\end{figure}

The promise of new, oxide-based electronic devices has been fueling the interest in and search for new multifunctional multiferroics (see Figures~\ref{fig1}(a) and b), a promising class of  materials that possess more than one ``ferro'' property such as ferroelectricity and (anti/ferro/ferri)magnetism. 
One of the biggest challenges in multiferroics research, which crosses many sub-disciplines of materials physics and chemistry, is finding or designing so-called cross-coupled multiferroics that operate at room-temperature. These are materials that are simultaneously magnetic and ferroelectric but in which the polarization (magnetization) can be manipulated in a useful way with an applied magnetic (electric) field, \textit{i.e.}, those multiferroics that display a magnetoelectric effect, as shown in Figure~\ref{fig1}(c).

Structurally and chemically complex oxides have highly tunable ground states and are thus strong candidate materials in which to realize new and enhanced magnetoelectric phenomena. The rewards could include not only new electronic device technologies, but new insights and challenges to our current understanding of the condensed matter sciences. The family of known oxides is very large, but there are also an enormous number of possible tertiary and quaternary materials that have yet to be identified, much less characterized. Recent advances in thin-film growth techniques mean that we can now also create materials that do not nominally exist in nature. Identifying promising candidate materials by surveying this vast space of design variables using a trial-and-error approach would be a hopeless task.

In the past five years, theorists have made rapid progress in the application of first-principles methods to the design of new multifunctional multiferroic materials. First-principles density functional methods have proved a powerful tool for studying the properties of materials at the level of atoms and electrons, without the need for empirical input. As such, the role of theory not only is that of analysis and interpretation of known experiments,  but -- increasingly -- its focus is on prediction~\cite{cohen86,cohen93,zunger99,spaldin03,fischer06,hafner06,ram12}. This has lead to more powerful ways of thinking about materials discovery where often theory is leading the way to the experimental discovery. In this article, we review the manner in which first-principles studies have increased our understanding of the magnetoelectric effect in transition metal oxides. We discuss the development of novel theoretical methods to calculate the magnetoelectric effect and the new insights gained from applying first-principles methods, all with a vision towards the rational design of new materials.

The organization of this review is as follows. In the remainder of the Introduction we discuss briefly the recent shift in our thinking about magnetoelectric design and a general strategy that has emerged. In
Section~\ref{lme} we focus the discussion on the linear magnetoelectric effect, not only the methodologies to calculate various contributions to it, but also the strategies to design materials with large linear magnetoelectric coupling are presented. In Section~\ref{beyondLME}, we discuss spin-lattice and
spin-phonon couplings and explain how they can be exploited to create new multiferroics and obtain divergent magnetoelectric responses. Microscopic coupling mechanisms and the effects of particular exchange-correlation functionals in calculations are also briefly discussed. Section~\ref{superlattice}
involves first principles approaches to artificial heterostructures, which are designed to display multiferroic properties.

Although our review is comprehensive, it cannot be exhaustive and we therefore decided to omit the topic of magnetically-induced ferroelectricity, which would include discussion on materials such as TbMnO$_3$. Many theoretical models have been proposed concerning the origin of magnetically induced ferroelectricity in antiferromagnetic oxides and materials have been designed. We refer the reader to the original papers and reviews and references therein.\cite{kimura03,arima11,sergienko06a,picozzi07,sergienko06b,katsura05,lee11}

Another important topic that we do not cover is the use of first-principles-based effective Hamiltonian approaches~\cite{zhong94}. First-principles calculations have been used for some time to obtain parameters for effective Hamiltonians, which are consecutively solved using Monte Carlo methods (see~\cite{zhong94, zhong96} for well-known examples). Although the use of such methods for the study of the magnotoelectric effect is relatively recent, it has been very successful. Some important examples include the calculation of not only linear but also higher-order magnetoelectric couplings in BiFeO$_3$ \cite{prosandeev11} and the prediction of the magnetotoroidic effect, where a curled electric field can switch the direction of the magnetization~\cite{ren11}. Again, we refer the reader to the original references for more information on this interesting topic~\cite{prosandeev11, ren11, sichuga10}. 

%

The basic method that most of the studies reviewed here uses is the Kohn-Sham formulation \cite{hohenberg64, kohn65} of  Density Functional Theory (DFT). This formulation uses electron density (as opposed to the full many body wave function) in order to calculate ground state properties of molecules or crystals, and thus decreases the computational cost of otherwise intractable problems significantly. As we do not aim a full review of DFT and related methods, we refer the reader to other reviews on basics of now standard DFT methods \cite{payne92, baroni01} and their applications to oxides and other materials \cite{rondinelli11}.

\subsection{Rethinking materials design }
Prior to the development of first-principles approaches to the magnetoelectric effect, symmetry arguments and a thermodynamic requirement were the primary guides for identifying materials displaying a large effect. The thermodynamic requirement states that the components of the {\it linear} magnetoelectric tensor, $\alpha$, should satisfy~\cite{brown68}
\begin{equation}
 \alpha_{ij} < \sqrt{\chi_{ii} \kappa_{jj}},
 \label{bound}
\end{equation}
where $\chi$ and $\kappa$ are the magnetic and dielectric susceptibilities respectively.  
This lead to the general belief that a large dielectric response, that is a large $\kappa$ or equivalently a large dielectric permittivity $\epsilon$, is required for a large linear ME effect, $\alpha$. Since ferroelectrics tend to be materials that display a large dielectric response (albeit in the paraelectric phase), a common misconception developed that magnetic-ferroelectrics, \textit{i.e.}, multiferroics, should be the targets of investigation for large linear magnetoelectric effects. 
By definition, multiferroics have both a spontaneous polarization and magnetization and therefore lack both space and time-reversal symmetries. In many cases, a linear ME effect would be allowed in such materials by symmetry. In the past, most design strategies have therefore focused on combining ferroic properties with the hope that a measurable ME response could be observed. Although this approach taught us much about new ferroic phenomena, it has enjoyed only modest success as means to identify materials for which there is a strong coupling between the various responses.
There are a few different reasons for this. First, note that the thermodynamic relationship expressed in Eq.~\ref{bound} is not a sufficient condition for strong coupling, that is, a large $\epsilon$ does not \emph{necessarily} lead to large $\alpha$, it is merely an upper bound. Furthermore, it has recently been pointed out that known ME materials have magnetoelectric responses that are far from pushing this limit~\cite{vanderbilt_private}. In short, just because symmetry allows an effect, it does \emph{not} follow that the effect will be strong.

With this in mind, the focus recently has shifted away from combining ferroic properties and moved in the challenging direction of understanding the fundamental mechanisms and key materials parameters that facilitate a strong cross-coupled response between the various ferroic orderings. 
In order to create targeted magnetoelectric phenomena, new approaches are being developed that begin with devising a mechanism that controls the interplay between the diverse microscopic degrees of freedom prevalent in these complex oxide ferroics. Researchers are then utilizing principles of crystal chemistry, solid-state physics, and symmetry to develop sets of design criteria to aid in the identification of a real  material with the physics of this mechanism built-in from the bottom-up. Finally, first-principles techniques, such as density-functional theory, are being used to screen for candidate materials and to incorporate materials-specific information into the conceptual models that ultimately guide the search for new materials showing unprecedented magnetoelectric responses.

\subsection{A general strategy for ``colossal'' responses }

Before moving on to the next section of this review it is worth discussing a general strategy that has been exploited in many different contexts to create multifunctional materials with large responses (and will be a common theme in much of what we later discuss): the idea of phase competition.
It has long been recognized that the key to enhancing sensitivity to an external perturbation is to position a material in the vicinity of multiple overlapping phase transitions (Bob Newhamm's Turnbull lecture on Smart Materials~\cite{newnham98} and Tokura's work on CMR manganites~\cite{tokura06} are classic papers). The morphotropic phase boundary in Pb(Zr,Ti)O$_3$ -  and the bicritical region in (La,Sr)MnO$_3$  are examples of regions where such overlapping transitions lead to highly reversible phase transformations easily traversed by an external field and subsequently to an increase in the piezoelectric and magnetoresistive performance of the material.  
The main point we wish to highlight is that although the colossal responses displayed by Pb(Zr,Ti)O$_3$ and (La,Sr)MnO$_3$ are categorically different phenomena, they share a common ``universal'' feature -- the competition among the respective microscopic degrees of freedom in each material leads to competing ordered states of matter.

In regards to the magnetoelectric effect, in 2006 it was argued~\cite{tokura07,fennie06b}  that the interplay between spins and polar lattice degrees of freedom can be exploited to produce a region of overlapping paraelectric-to-ferroelectric and antiferromagnetic-to-ferromagnetic phase transitions. 
The important thing about this strategy is that that the competition between these different ferroic orders in this Òphase competitionÓ region of the phase diagram, as depicted in Figure~\ref{fig1}(d), could lead to effectively ``colossal'' magneto-electric and magneto-capacitive responses, without the need for the microscopic magnetoelectric coupling itself to be large. Additionally, it opens a route to achieve a large, non-trivial ME response from microscopic mechanisms beyond the linear ME effect, \textit{i.e.}, at  all orders of coupling, which removes many, if not all of the symmetry restrictions imposed by the linear effect.

The  basic idea behind these design strategies to create strongly-coupled magnetic-ferroelectrics may seem counter-intuitive when considering the past obsession with multiferroics. Here the starting point can be in fact an antiferromagnetic-paraelectric (AFM-PE) material, which is the phase most magnetically ordered insulators have as their ground state.  The challenge is two-fold: first, identify such AFM-PE materials that have a ferroelectric-ferromagnetic metastable phase (or in a previously inaccessible region of the phase diagram); and second, identify an external control parameter that can be exploited to tune between the various phases. Here first-principles approaches have a distinct advantage at mapping out this huge phase space that is difficult, if not impossible, to access experimentally -- {\it First principles to the rescue!}

\section{\label{lme}The Linear Magnetoelectric Effect}


A discussion on  magnetoelectricity in transition metal oxides inevitably starts with the linear magnetoelectric  (ME) effect, a topic that has received much attention over the years \cite{fiebig05, fiebig09}. This linear effect involves the induction of an electric polarization that is proportional to an external magnetic field $H$, and likewise, a magnetization that is proportional to an external electric field $E$. Pierre Curie was the first to predict the existence of such an effect over one hundred years ago;~\cite{curie94} the first experimental observations were made almost 50 years ago~\cite{astrov60}. 
Phenomenologically, it is related to the lowest order (bilinear) coupling between $E$ and $H$ in the free energy \cite{landau84}: 
\begin{equation}
\mathcal{F}=\mathcal{F}_0-\frac{1}{2}\kappa_{ij}E_iE_j-\frac{1}{2}\chi_{ij}H_iH_j-\alpha_{ij}E_i H_j,
\label{eq1}
\end{equation}
where Latin indices represent spatial directions. $\kappa$ and $\chi$ are the dielectric and magnetic susceptibilities, and $\alpha$ is the linear magnetoelectric tensor. $\alpha$ is an axial tensor of second rank that is subject to very strict symmetry constraints~\cite{borovik-romanov06}: only 58 of the 122 magnetic point groups allow a nonzero linear magnetoelectric effect. In addition, except for 2 of these groups, multiple components of $\alpha$ are set to zero by symmetry (a detailed analysis of the form of $\alpha$ for all magnetic point groups can be found in~\cite{borovik-romanov06}). Given these rather stringent symmetry restrictions, the fact that the linear magnetoelectric effect was predicted and subsequently experimentally confirmed to exist in a number of materials is a significant accomplishment.

Despite these early achievements, there is no known material that displays a large linear magnetoelectric effect at room temperature. 
However, in recent years new first-principles methodologies have made it possible to calculate the linear magnetoelectric tensor in a relatively straightforward manner and in some cases with little or no modification of existing first-principles codes. 

Theory plays an important role in the search for new magnetoelectric materials by guiding experimentalists in their investigations and by helping to screen promising candidate materials.  In this section we discuss new first-principles methods developed to calculate the different microscopic contributions to the linear ME effect, displayed in Figure~\ref{fig:linearMETable}, and the insights gained from the application of these methods.


\subsection{Advances in calculating the linear magnetoelectric tensor from first principles}

\begin{figure}
  \begin{center}
    \includegraphics[width=1.0\hsize]{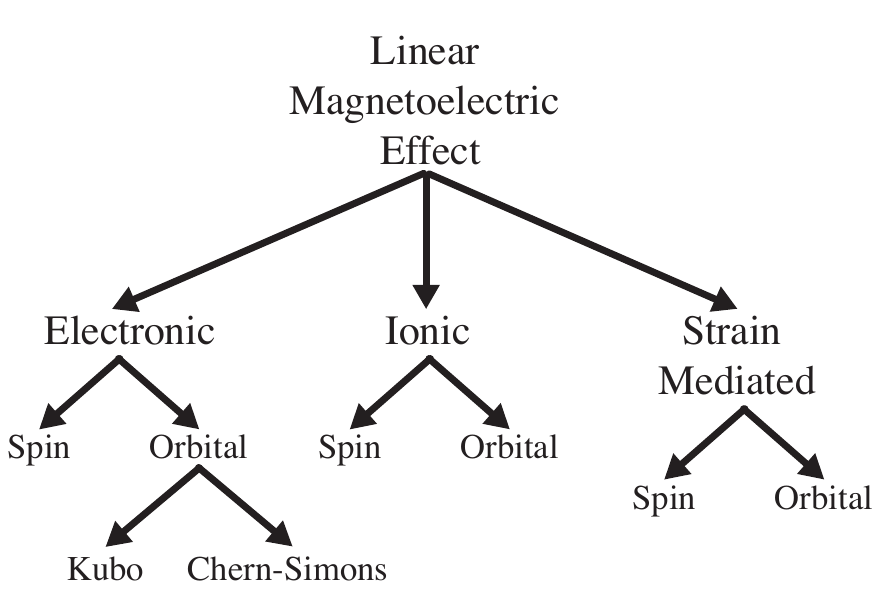}
  \end{center}
    \caption{Different contributions to the linear magnetoelectric effect at zero temperature. }
    \label{fig:linearMETable}
\end{figure}

\begin{figure}
  \begin{center}
    \includegraphics[width=1.0\hsize]{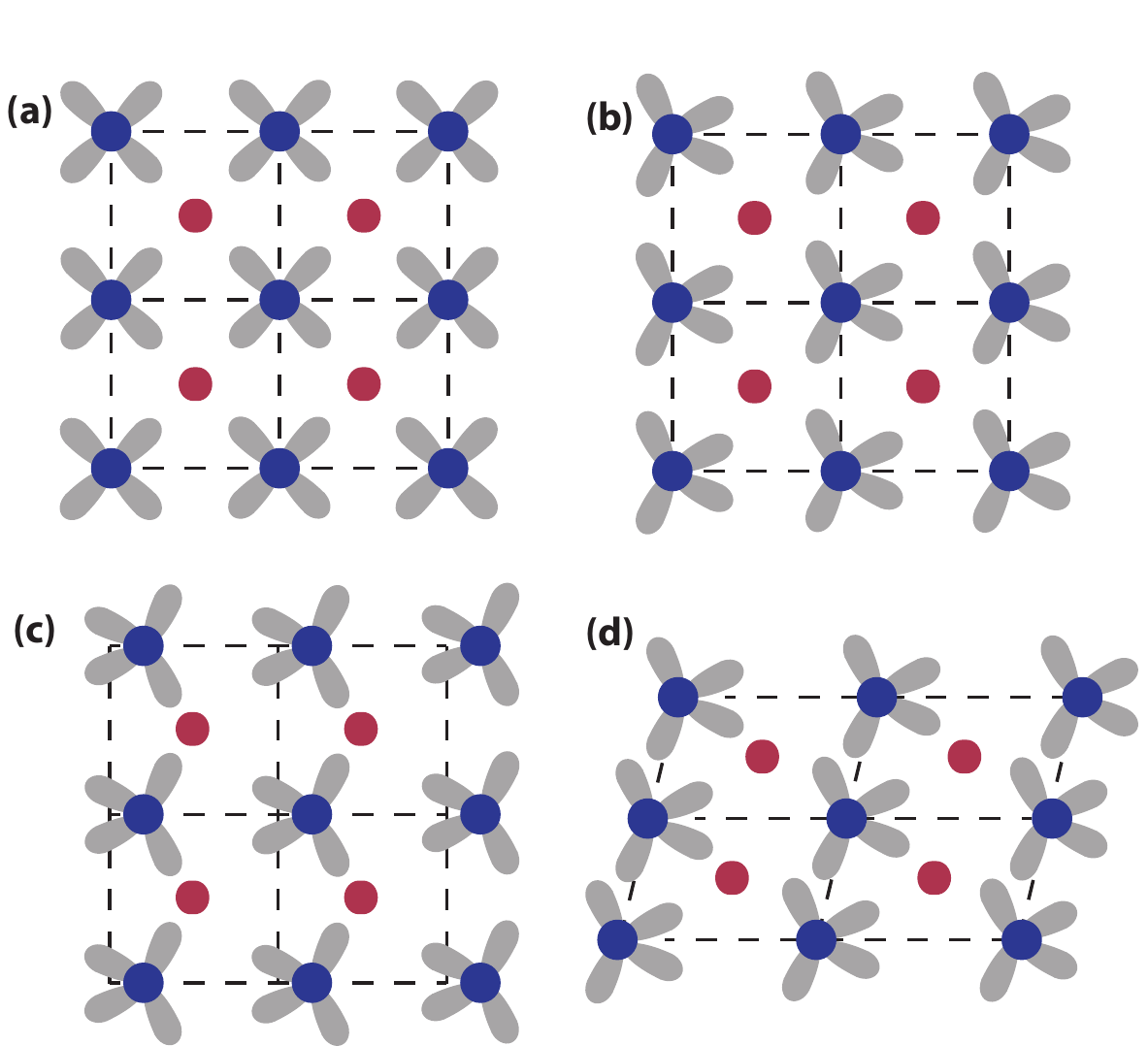}
  \end{center}
    \caption{Sketch for different contributions to spin part of the linear magnetoelectric tensor: (a) The high symmetry system under no external field. (b) The clamped-ion response to an external field, $\alpha^\mathrm{elec}$, where the electron wavefunctions have changed but not the ionic positions. (c) The change in ionic positions due an external field, accompanied by a change in the electron wavefunctions due to this change. This contribution corresponds to $\alpha^\mathrm{ion}$. (d) Lattice vectors can change in response to an external field, giving rise to $\alpha^\mathrm{strain}$. }
    \label{fig:linearME}
\end{figure}

The linear magnetoelectric tensor, $\alpha$, can be written as the sum of
three terms:
\begin{equation}
\alpha=\alpha^\mathrm{elec}+\alpha^\mathrm{ion}+\alpha^\mathrm{strain},
\end{equation}
where $\alpha^{\mathrm{elec}}$ is the purely electronic part,
$\alpha^{\mathrm{ion}}$ is the ionic contribution, and
$\alpha^{\mathrm{strain}}$ is the strain-mediated magnetoelectric response
tensor, as shown in Figure~\ref{fig:linearMETable}.
The electronic contribution  arises from the change in the electronic polarization under an external magnetic field, or equivalently from the change in magnetization due to the change in electronic wave function under an external electric field. 
The ions are assumed to be fixed in their positions, as shown in Fig. \ref{fig:linearME}(b). This clamped ion response can be measured at high frequencies where the ions cannot respond fast enough to changes in the external field.
Both $\alpha^{ion}$ and $\alpha^\mathrm{strain}$ are lattice mediated, where the former is related to the change in magnetization/polarization with respect to the internal ionic positions (Fig. \ref{fig:linearME}(c)) and $\alpha^\mathrm{strain}$ arises from changes in the unit cell shape under external fields and so is related to partial derivatives of the magnetization/polarization with respect to the unit cell vectors (Fig. \ref{fig:linearME}(d)). 
Each of the three contributions to $\alpha$ can further be decomposed into spin and orbital magnetization parts. We
start by reviewing the studies that restricted themselves to the spin
contribution as practical matter.


\subsubsection{Spin-only magnetization at T=0}
The development of first-principles methodologies to quantify $\alpha$ (or its various contributions) was started in 2008 by \'{I}\~{n}iguez~\cite{iniguez08}. In this work, $\alpha^\mathrm{elec}$ was assumed to be negligible and only $\alpha^\mathrm{ion}$ was considered. It was shown that $\alpha^{ion}$ has a very elegant form in terms of the Born mode effective charges, $Z^e$, defined as the derivative of polarization, $P$, with respect to a phonon-mode amplitude, $u$, \textit{i.e.},
\begin{equation}
Z^e_{ni}=\Omega_0\frac{\partial P_i}{\partial u_n},
\end{equation}
where $P_i$ is the $i^{th}$ component of the polarization, $u_n$ is the amplitude of the $n^{th}$ normal mode and $\Omega_0$ is the unit cell volume. 
In the past, there was serious debate \cite{cochran61} on how to define atomic charges in a solid. The Born mode effective charge (and more generally the Born Effective Charge~\cite{ghosez98}) is well-defined, experimentally measurable~\cite{axe71}, and has played a central role in the theory of dielectrics.
For example, the ionic contribution to the dielectric tensor has a very simple expression in terms of these mode effective charges \cite{cochran62}: 
\begin{equation}
\epsilon^{ion}_{ij}=\frac{4 \pi e^2}{\Omega_0} \sum_{n=1}^{N_{IR}}  \frac{Z^e_{ni}Z^e_{nj}}{\omega^2_n}
\label{equ:epsilon}
\end{equation}
where the summation runs over the infrared-active modes, $e$ is the elementary charge and $\omega_n$ is the frequency of the $n^{th}$ mode. 

What \'{I}\~{n}iguez did is to define a magnetic analogue of the mode effective charge as the derivative of magnetization with respect to mode amplitude: 
\begin{equation}
Z^m_{ni}=\Omega_0\frac{\partial M_i}{\partial u_n},
\end{equation}
where $M_i$ is the $i^{th}$ component of magnetization. Using this definition and considering Eq.~\ref{eq1}, it is possible to obtain an elegant formula for $\alpha^{ion}$,
\begin{equation}
\alpha^\mathrm{ion}_{ij}=\frac{1}{\Omega_0}\sum_{n=1}^{N_{IR}}\frac{Z_{ni}^e Z_{nj}^m}{K_n^2},
\label{equ:alpha_ion}
\end{equation} 
which has an aesthetically pleasing (and intuitive) form similar to Eq.~\ref{equ:epsilon}. (Here, $K_n$ are the eigenvalues of force constant matrix.) 
Both $Z^e$ and $Z^m$ can be calculated by finite differences using standard DFT codes, so the expression for the ionic contribution to the linear magnetoelectric effect can now be  calculated in a straightforward manner for systems of interest. 
Although \'{I}\~{n}iguez calculated only the spin contribution to $\alpha$, this method can also be used to calculate the orbital contribution, as has been done recently.\cite{scaramucci12}

This formalism was extended to the strain-mediated magnetoelectric response $\alpha^\mathrm{strain}$ in Refs. \cite{wojdel09} and \cite{wojdel10}. This contribution to $\alpha$ comes from a combination of piezoelectric and piezomagnetic effects and can be expressed in terms of the piezoelectric tensor $e$, the piezomagnetic tensor $h$, and the elastic tensor $C$ as 
\begin{equation}
\alpha^{strain}=eC^{-1}h.
\label{equ:alpha_strain}
\end{equation} 
The intuitive meaning of this expression is that, for example, an external magnetic field leads to a strain proportional to $C^{-1}h$, which in turn leads to a piezoelectric polarization proportional to $eC^{-1}h$. So the resultant coupling between the electric and magnetic fields is proportional to the strengths of both piezoelectric and piezomagnetic couplings. Note that all three tensors on the right hand side of Equation \ref{equ:alpha_strain} can be calculated from first principles by finite differences. Hence, standard DFT codes can be used to calculate $\alpha^\mathrm{strain}$ as well. The orbital contribution to $\alpha^{strain}$ can be evaluated by this formalism too, but no such calculation has been presented yet to the best of our knowledge. 

Bousquet, Spaldin and Delaney calculated $\alpha^\mathrm{elec}$ from first-principles for the first time and challenged the assumption that the purely electronic magnetoelectric response is small compared to the ionic contribution~\cite{bousquet11a}. Although it is necessary to use a full vector potential to simulate an applied magnetic field properly, the authors chose a computationally easier approach and considered only a Zeeman field as an approximation. The sole effect of the Zeeman field is to split the spin-up and spin-down states of electrons. This approach requires only minimal modification of existing publicly available DFT codes. In Ref.~\cite{bousquet11a}, this is done for the VASP code~\cite{kresse96} and it was shown that $\alpha^\mathrm{elec}$ is comparable to $\alpha^\mathrm{ion}$ in Cr$_2$O$_3$.

As discussed so far, the linear magnetoelectric effect requires an interaction between the charge and spin distributions of the material, in other words, between the orbital and spin degrees of freedom of the electrons. This interaction arises from spin-orbit coupling (SOC), which is a relativistic effect ~\cite{yosida01, cohen-tannoudji92}. In the reference frame of the electron, the nucleus orbits around it and hence creates a magnetic field that is proportional to the orbital angular momentum of the electron. This leads to an energy term that is proportional to $\vec{L}\cdot\vec{S}$, where $\vec{L}$ and $\vec{S}$ are the orbital and spin angular momenta of the electron. This SOC term connects the orbital and spin degrees of freedom and breaks the spin rotation symmetry for the symmetries of the lattice. Not only is SOC responsible for phenomena like magnetocrystalline anisotropy and the Dzyaloshinskii - Moriya interaction (discussed in appendix), there would be no linear ME effect at $T=0$ without spin-orbit coupling.

\subsubsection{Spin-only magnetization at finite temperature}
If the temperature is not zero, a different contribution to the linear ME effect emerges. Mostovoy and co-workers studied the effect of spin fluctuations in Cr$_2$O$_3$ at finite temperatures using a combination of DFT calculations and Monte Carlo simulations~\cite{mostovoy10}. The groundstate of Cr$_2$O$_3$ has no electric polarization because it is forbidden by symmetry (the point-group operation ``space inversion followed by time reversal'' leaves the system intact). At nonzero temperatures, however, it is possible that spin fluctuations  break the relevant symmetries and lead to an electric polarization. An external magnetic field couples to the spin fluctuations and hence to the electric polarization. This leads to an extra contribution to $\alpha$ that exists only at nonzero temperature. An important difference of this contribution from the previously described zero-temperature effects is that it does not rely on spin-orbit coupling but originates solely from the Fermi statistics and nonrelativistic Heisenberg exchange interactions. In the well-studied material Cr$_2$O$_3$, Mostovoy, \textit{et al.,} found that the linear magnetoelectric effect originating from the relativistic effects is about an order of magnitude smaller than the finite temperature contribution~\cite{mostovoy10}. This suggests that the search for large $\alpha$ materials should be extended to non-zero temperatures as well.

\subsubsection{Orbital magnetization}

All of the studies mentioned in this section so far consider only the spin magnetization of electrons. However, as charged particles in motion, electrons in a solid can also have an orbital magnetization and a corresponding magnetoelectric response~\cite{OMEP1, OMEP2, OMEP3, OMEP4}. The orbital contribution to magnetization had not been considered in first-principles calculations until the recent developments in the modern theory of magnetization~\cite{MTMReview, MTM1, MTM2, MTM3}. The clamped-ion orbital magnetoelectric response tensor can be cast in a form that is the sum of a \textit{Kubo} part and a pseudoscalar \textit{Chern-Simons} part, which is isotropic. The Chern-Simons contribution can be written as,
\begin{equation}
\alpha^{CS}_{ij}=\theta\frac{e^2}{2\pi h}\delta_{ij},
\label{alpha_cs}
\end{equation}
where $e$ is the elementary charge, $h$ is Planck's constant and $\delta_{ij}$ is the Kronecker delta. $\theta$ is a dimensionless parameter (defined modulo $2\pi$), which originally appeared in the context of axion electrodynamics\cite{wilczek87}. Axions are hypothetical elementary particles which were introduced to explain charge-parity violation \cite{franz08}. If they were to exist, the Maxwell equations would be modified in a way that would lead to a bilinear coupling between electric and magnetic fields. While they don't exist as elementary particles in free space, axions have attracted a lot of attention in the field of topological insulators \cite{hasan10}. It is an interesting twist of fate that the idea of axions, originally conceived in the discipline of high-energy physics, is finding widespread application in the field of condensed matter physics. 

Even though $\alpha^{CS}$ has a compact form, there are many subtle details involved in the calculation of the orbital magnetization (and therefore $\alpha^{CS}$) of an extended solid, which can be circumvented by the use of Wannier functions~\cite{wannier37, marzari97} instead of Bloch states in calculations. Coh and co-workers performed a detailed theoretical analysis of $\alpha^{CS}$ using this Wannier function approach~\cite{coh11}.

While we were preparing this
review a pre-print by Scaramucci and co-workers \cite{scaramucci12} appeared on the
ion-mediated orbital contribution to $\alpha$. The first calculation of
the Kubo terms for a real material has recently been presented by Malashevich, et al.\cite{malashevich12} 

\subsection{The power of first-principles methodologies: Elucidating new design strategies for achieving a large linear magnetoelectric effect}

Developments in the understanding and calculation of different contributions to the linear magnetoelectric effect reviewed in the previous section have lead to efforts to rationally design materials displaying a large linear magnetoelectric effect.

Here we will see examples of the general theme discussed in the Introduction to produce large responses. Specifically, strategies were developed to design materials with a strong linear ME effect for materials in which $\alpha$ is in fact  relatively small and even zero by symmetry in the bulk ground state. The novel idea was to identify an external control parameter to tune the system to the proximity of a structural phase transition.

\subsubsection{Enhancing the linear ME effect}

\begin{figure}
\centering
\includegraphics[width=8cm]{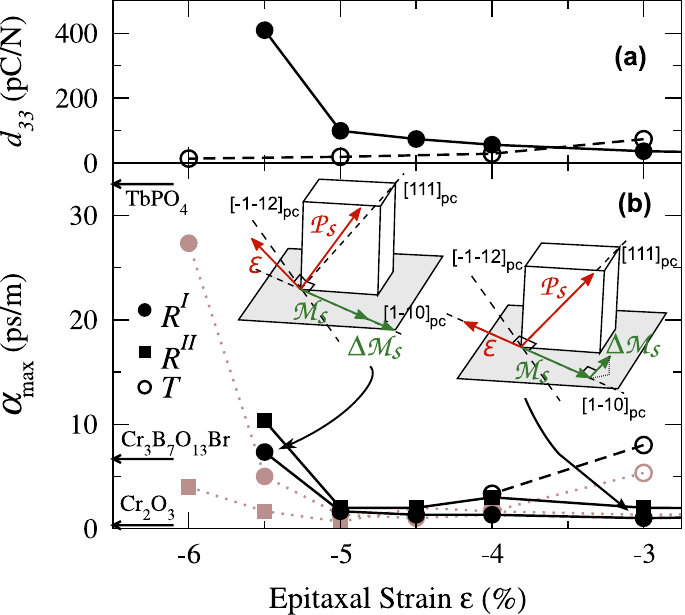}
\caption{First principles calculations of structural softness leading to an enhanced a) piezoelectric and b) magnetoelectric response in BiFeO$_3$. 
The transition between the $R$ and $T$ phases occurs at 4.4\% strain, with hysteresis extending from -6\% to -3\% strain. The frozen-cell (light) and full lattice-mediated (dark) response are enhanced in both the R and T phases close to the boundary of this hysteresis region. $\alpha_{max}$ is the largest component of $\alpha$. A sketch of $\alpha_{max}$ is provided in the inset for the $R$ phase. 
From \cite{wojdel10}. Copyright (2010) by the American Physical Society.}
\label{wojdel10-figure3}
\end{figure}

Wojde\l~and \'{I}\~{n}iguez used the idea of `structural softness' in their first-principles study of magnetoelectricity in strained BiFeO$_3$~\cite{wojdel10}. They defined a material as structurally soft if its force-constants matrix ($K$) or elastic tensor ($C$) had a vanishingly small eigenvalue. Noting that the lattice-mediated contributions to $\alpha$ depend on $K$ and $C$, they argued that structurally soft materials should have large values of $\alpha$. This can be seen from Eq. (\ref{equ:alpha_ion}), which indicates that $\alpha^{ion}$ diverges as $K_n$ approaches zero. Because $\epsilon^{ion}$ also diverges, positioning a system close to a ferroelectric phase transition should result in the desired enhancement. 
Selecting BiFeO$_3$ as their test material, Wojde\l~and \'{I}\~{n}iguez calculated and compared the stability, piezoelectric and magnetoelectric responses of two ferroelectric phases, namely the rhombohedral ($R$) and tetragonal ($T$) phases, as a function of epitaxial strain. They found that for the $R$ phase in particular, large compressive strains ($\sim$6\%) induced a large increase in $\alpha$, as shown in Figure \ref{wojdel10-figure3}. The authors argued that the enhancement is associated with a phonon mode with a vanishing force-constant. Ideally, the soft mode associated with the enhancement of $\alpha$ should have a large mode effective electric charge $Z^e$ and magnetic `charge' $Z^m$. As noted by Wojde\l~and \'{I}\~{n}iguez, this is not the case with BiFeO$_3$, but a significant enhancement is observed nonetheless.

\subsubsection{Inducing the linear ME effect or weak ferromagnetism: Two sides of the same coin }

First-principles calculations also recently predicted a large magnetoelectric effect in the antiferromagnetic insulator CaMnO$_3$~\cite{bhattacharjee09,bousquet11b}. CaMnO$_3$ adopts the orthorhombic $Pnma$ structure in its ground state but first-principles calculations~\cite{bhattacharjee09} revealed the existence of a (weak) zone-center ferroelectric instability in the cubic $Pm\bar{3}m$ phase, which is suppressed by the octahedral rotations in the ground state. Bhattacharjee and co-workers have shown however that a proper ferroelectric instability could be induced by tensile strain. Additionally, they showed that the ferroelectric instability is Mn-dominated, contrary to earlier arguments~\cite{filippetti02} that displacement of the B-site cation from its ideal $Pm\bar{3}m$ position would not be favored in materials with partially filled $d$ orbitals. \cite{bersuker12} Hence, Bhatacharjee, \textit{et al.,} demonstrated that magnetic ions could contribute to ferroelectric distortions, an important development in the search for multiferroics. 

In 2011, Bousquet and Spaldin~\cite{bousquet11b} showed that strain can induce a linear magnetoelectric effect in CaMnO$_3$ by inducing ferroelectricity.
In $Pnma$ CaMnO$_3$ symmetry allows a small canting of the antiferromagnetic spins, that is, weak-ferromagnetism, but forbids a linear magnetoelectric effect (a modest effort can be used to show that in a material with a center of inversion, the linear magnetoelectric effect and weak-ferromagnetism are mutually exclusive; see the Appendix).
Using first-principles calculations, Bousquet and Spaldin showed that $Pnma$ CaMnO$_3$ can be driven ferroelectric with strain. Since a ferroelectric distortion removes the symmetry that forbids the existence of the linear magnetoelectric effect in the paraelectric structure, the linear magnetoelectric tensor is proportional to the ferroelectric polarization and  therefore changes in $\alpha$ can be quite large (note that since  the ferroelectric distortion may in fact change the type of collinear, antiferromagnetic order, this conclusion of ferroelectrically-induced linear magnetoelectricity is only true if in the ferroelectric $Pmc2_1$ structure,  the collinear spin structure remains that which allows a linear magnetoelectric effect; Bousquet and Spaldin explicitly confirmed this for CaMnO$_3$).
The symmetry analysis presented in Ref.~\cite{bousquet11b} is valid for all $Pnma$ ABX$_3$ perovskites. Considering that $Pnma$ is the most common space group among the ABO$_3$ perovskites, Bousquet and Spaldin's approach represents a promising route to search for new magneotelectrics.



In 2008, Fennie~\cite{fennie08,ederer08} formulated a set of  structural and magnetic criteria that must be satisfied in a particular material in order for a ferroelectric distortion to induce weak ferromagnetism, a idea first suggested back in 1977 by Fox and Scott~\cite{scott77}. Quite simply, the criteria are the complement of the ideas expressed by Bousquet and Spaldin in 2011. Fennie argued that the paraelectric-antiferromagnetic reference structure should allow a linear magnetoelectric effect, which subsequently forbids weak ferromagnetism. Then, the ferroelectric distortion could induce weak ferromagnetism and as such the induced magnetization would be proportional to the ferroelectric polarization.
Using FeTiO$_3$ as the test case, Fennie then used first-principles calculations to show that both criteria are satisfied in a series of $R3c$ ATiO$_3$ materials (A=Fe, Ni, Mn). Experiments subsequently verified that $R3c$ FeTiO$_3$ is both ferroelectric and a weak ferromagnet~\cite{varga09}. These predictions sparked a search for other materials satisfying the same criteria, a number of which have also been recently synthesized~\cite{inaguma08,son09,aimi11}.

The work of Fennie and that of Bousquet and Spaldin showed the utility of using a somewhat ``boring'' paraelectric reference structure to search for ``exciting'' phenomena in ferroelectric distorted ground states. We point out that the symmetry in the ground state of the materials considered by the different authors allows ferroelectricity, weak-ferromagnetism and the linear magnetoelectric effect. However, qualitatively different responses to applied strain occur in each material, which becomes clear when using a higher-symmetry reference-structure. For example, increasing the polarization in strained CaMnO$_3$ would, to lowest order, increase the linear magnetoelectric effect while not increasing the net magnetization, whereas the exact opposite would occur for FeTiO$_3$ (in the Appendix we provide a simple Landau model to show this).


We have discussed an example where a ferroelectric distortion induces the linear magnetoelectric effect and an example where a ferroelectric distortion induces weak ferromagnetism. Can a single material display both effects? What we mean is, can a ferroelectric distortion induce both  the linear magnetoelectric effect and weak-ferromagnetism in a structure that displays neither?
In order for this to happen, the ferroelectric distortion must change the translational symmetry of the material, which is not possible in a proper ferroelectric transition like that in strained CaMnO$_3$ and FeTiO$_3$. It  could occur however in an improper ferroelectric transition. Recently, Benedek and Fennie~\cite{benedek11} have discovered such a material: the $n=2$ Ruddlesden-Popper material Ca$_3$Mn$_2$O$_7$. They showed that the ferroelectric distortion was driven by two different octahedral rotation patterns with different symmetries~\cite{benedek11}. Specifically, there is a trilinear term in the free energy of the form,
\begin{equation}
\mathcal{F}=\gamma P R_1 R_2,
\end{equation}
where $P$ is the polarization and $R_1$ and $R_2$ are the octahedral rotations. 
The combined rotation pattern $R_1 \oplus R_2$ is treated as a `hybrid' distortion and is what drives the system into the polar state, that is, the polarization becomes non-zero only when both rotations condense~\cite{harris11}. For these reasons, Benedek and Fennie coined the term hybrid improper ferroelectricity to describe this mechanism, which was first discovered by Bousquet, Dawber and co-workers~\cite{bousquet08} in SrTiO$_3$/PbTiO$_3$ superlattices.
(Also see Ref. \cite{fukushima11} for the same effect in double perovskites.) 
In the case of Ca$_3$Mn$_2$O$_7$, the octahedral rotations induce not only ferroelectricity, but also weak ferromagnetism and a linear magnetoelectric effect. 
Rondinelli and Fennie subsequently elucidated the design criteria for the creation of new hybrid improper ferrroelectrics using non-polar $Pnma$ perovskites as `building blocks'~\cite{rondinelli12}. As mentioned previously, $Pnma$ is the most commonly adopted space group among ABO$_3$ perovskites~\cite{lufaso01}, hence the study and synthesis of hybrid improper ferroelectrics represents a very promising route to the discovery of new magnetoelectric materials. The implications of rotation-driven ferroelectricity for the design of multifunctional materials were recently discussed in Ref.~\cite{benedek12}.

\subsubsection{Inducing the linear ME effect in a topological insulator }

Although  $\alpha^{CS}$ is much smaller than the spin counterpart for the conventional magnetoelectrics considered so far, Coh, \textit{et al.,} argued that it may be possible to obtain very large orbital magnetoelectric couplings in materials close to a topological insulator \cite{hasan10} state.
$\theta$ is usually small in ordinary insulators, but it is equal to $\pi$ in topological insulators (TI). This might lead one to think that TIs would exhibit a large linear ME, however the gapless surface states in TIs cancel the magnetoelectric contribution from the bulk and hence the total magnetoelectric response is zero~\cite{coh11}. Note that it is time reversal symmetry that \textit{topologically protects} these surface states, and equivalently, sets $\alpha$ to zero. However, one can imagine a material that does not have time reversal symmetry but is close to a TI state. Such a state can be obtained by breaking the time reversal symmetry of a TI by a small perturbation. This would lead to a large but nonquantized $\theta$, namely $\theta \sim \pi$. There are no topologically protected surface states in such a material so the linear ME response from the bulk can be observed. Coh and co-workers~\cite{coh11} demonstrated the validity of this idea by breaking the time reversal symmetry in the well-studied TI Bi$_2$Se$_3$ by a fictitious Zeeman field and calculated $\theta$ from first principles. While this provides a proof of principle, discovering a material with strong orbital magnetoelectric response is currently one of the grand challenges in the field.

\section{Beyond Linear Magnetoelectric Effect: Spin -- Phonon/Lattice Coupling}
\label{beyondLME}
The linear magnetoelectric effect is the lowest order coupling between the electric and magnetic degrees of freedom, however, it is not the only possible coupling. Given the rather restrictive symmetry requirements of the linear effect, researches have sought ways to achieve nontrivial magnetoelectric effects from higher-order coupling of  the electric and magnetic fields. One such higher-order coupling that has played a central role in the search for magnetoelectrics in recent years is the biquadratic coupling, $F\sim E^2H^2$, which is allowed by symmetry in any magnetic point group.

One reason this term has been examined is because there is a simple physical origin of the term, which can be used as a guide to search for new materials and new effects. The microscopic mechanism stems from the coupling of spins to the zone center, polar force-constants, often approximated in the literature by the following expression~\cite{baltensperger70}
\begin{equation}
\omega \approx \omega_{PM} +\gamma\langle\mathbf{S}_i\cdot\mathbf{S}_j\rangle,
\label{spc-eq1}
\end{equation}
where $\omega_{PM}$ is the phonon frequency in the paramagnetic state well above a magnetic ordering temperature where local spins have formed but can be considered as uncorrelated, $\gamma$ is the spin-phonon coupling constant, and $\langle\mathbf{S}_i\cdot\mathbf{S}_j\rangle$ is the nearest-neighbor spin-spin correlation function (note that in general the phonon frequency may also depend on correlation functions between further neighbors). This spin-phonon coupling exists in magnetic materials as a consequence of the fact that the magnetic exchange parameters depend on ionic positions \cite{baltensperger70}. In this section, we will give a brief introduction to the microscopic mechanism, discuss advances in our understanding of spin-phonon coupling in a few example materials and discuss the applicability of first-principles methods. Additionally, we will see how spin-lattice coupling, that is, the interdependence of magnetic and structural orderings (which also stems from higher-order couplings) has been used to create novel magnetoelectric responses. 

\subsection{Spin-Phonon and Spin Lattice coupling}
We begin with a few words regarding nomenclature. In principle, spin-phonon coupling refers to a dynamic effect in which spin-correlations influence phonon frequencies and/or vice versa. Lawes and co-workers performed an elegant set of experiments demonstrating how such coupling leads to distinct dielectric constant vs temperature behavior in antiferromagnetic and ferromagnetic materials~\cite{lawes03}. The manner in which magnetic correlations influence the dielectric behavior of an insulator follows from examination of the expression for the ionic contribution to the dielectric constant, Eq.~\ref{equ:epsilon}, which goes like the inverse phonon frequency squared. Therefore, intrinsic magnetopermittivity has its origin in spin-phonon coupling. However, most studies on spin-phonon coupling in the field of multiferroics are really discussing spin-lattice coupling: the effect of the magnetic order/correlations on the force constant, an inherently static quantity. This misnomer carried over from the field of ferroelectrics, in particular the soft-mode theory of ferroelectricity in which a soft polar ``phonon'' is central to the theory (in reality, it is a softening of the lattice, that is, the force constant). We won't discuss these subtleties any further but unless otherwise noted we are discussing the coupling of spins to the force constants of the crystal.  


Techniques for the \emph{ab initio} calculation of phonon frequencies and magnetic exchange parameters are well developed but researchers have only recently started calculating spin-phonon coupling~\cite{fennie06, sabiryanov99}. The starting point has been the following model for the total energy of the system:
\begin{equation}
E = \frac{1}{2}\sum_{nm\alpha\beta}C^{\alpha\beta}_{nm}u_{n\alpha}u_{m\beta} - \sum_{ij}J_{ij}\mathbf{S}_i\cdot\mathbf{S}_j + E_0,
\label{spc-eq2}
\end{equation}
where the first term is the internal elastic energy with $C^{\alpha\beta}_{nm}$ as a bare force constant matrix, that is, the force constants in the paramagnetic state (note, this is not the same as the non-magnetic state, here we mean that local moments have formed but are uncorrelated) and $u_{n\alpha}$ denotes the displacement of atom $n$ from an equilibrium position in the Cartesian direction $\alpha$. The second term is the magnetic energy represented by a Heisenberg model (a  good approximation for a large class of magnetic insulators) with $\mathbf{S}_i$ as the spin of magnetic ion $i$ and $J_{ij}$ as the exchange parameters. Finally, $E_0$ is that part of the energy that doesn't depend on spins or atomic displacements. 

The  idea (introduced decades ago~\cite{baltensperger70}) which accounts for spin-phonon coupling is that in general the $J_{ij}$'s not only depend on the positions of magnetic ions but also on the positions of the surrounding nonmagnetic atoms such, as the neighboring oxygen atoms. For small displacements, we can expand $J_{ij}$ in a Taylor series up to second order in the displacements, for which it has been recently suggested~\cite{fennie06} that the expansion should be done in the symmetrized basis functions, $\eta$, 
\begin{equation}
J_{ij} = J^{0}_{ij} + \frac{1}{2}\sum_{nm\alpha\beta}\frac{\partial^2J_{ij}}{\partial \eta_{n\alpha}\partial \eta_{m\beta}}\eta_{n\alpha}\eta_{m\beta},
\label{spc-eq3}
\end{equation}
rather than in the ionic positions of two spins and angle {\it w.r.t.}~the connecting anion. Here, $n$ and $m$ label irreducible representations and $\alpha$ and $\beta$ label directions. $J^{0}_{ij}$ are exchange parameters for atoms in their equilibrium positions. Here the assumption is that the first-order term is zero, which is true unless the crystal is a linear magnetoelectric (see Section \ref{lme}). By substituting (\ref{spc-eq3}) into (\ref{spc-eq2}) and integrating over the magnetic degrees of freedom one finds that the elastic energy is given by $\frac{1}{2}\sum_{nm\alpha\beta}\tilde{C}^{\alpha\beta}_{nm}u_{n\alpha}u_{m\beta}$ with the renormalized force-constant matrix \cite{fennie06}
\begin{equation}
\tilde{C}^{\alpha\beta}_{nm} = C^{\alpha\beta}_{nm} - \sum_{ij}\frac{\partial^2J_{ij}}{\partial \eta_{n\alpha}\partial \eta_{m\beta}}\langle\mathbf{S}_i\cdot\mathbf{S}_j\rangle
\label{spc-eq4}
\end{equation}
\begin{equation}
\Rightarrow \tilde{\omega}^2 \approx \omega_{\rm PM}^2 - \lambda\langle\mathbf{S}\cdot\mathbf{S}\rangle
\label{freq}
\end{equation}
where $\omega$ is the phonon frequency.  Since the bare force constants are not accessible from a static theory such as density functional theory, $C^{\alpha\beta}_{nm}$ cannot be directly calculated. Instead, Ref.~\cite{fennie06} suggested that one should evaluate the force-constant matrix for several ordered, low-energy magnetic configurations (i.e., the left hand side of Eq.~\ref{spc-eq4}). This allows one to obtain the derivatives $\frac{\partial^2J_{ij}}{\partial \eta_{n\alpha}\partial \eta_{m\beta}}$ by fitting them to Eq.~\ref{spc-eq4} (since $\langle\mathbf{S}_i\cdot\mathbf{S}_j\rangle$ is known for the ordered system), analogous to how the magnetic exchange parameters, $J_{ij}$, are calculated using total-energy calculations. Spin-phonon coupling parameters can then be found as demonstrated for the spinel ZnCr$_2$O$_4$~\cite{fennie06}.

\subsection{EuTiO$_3$ and the idea of phase competition}

\begin{figure}[t]
\begin{center}
\includegraphics[width=0.5\textwidth]{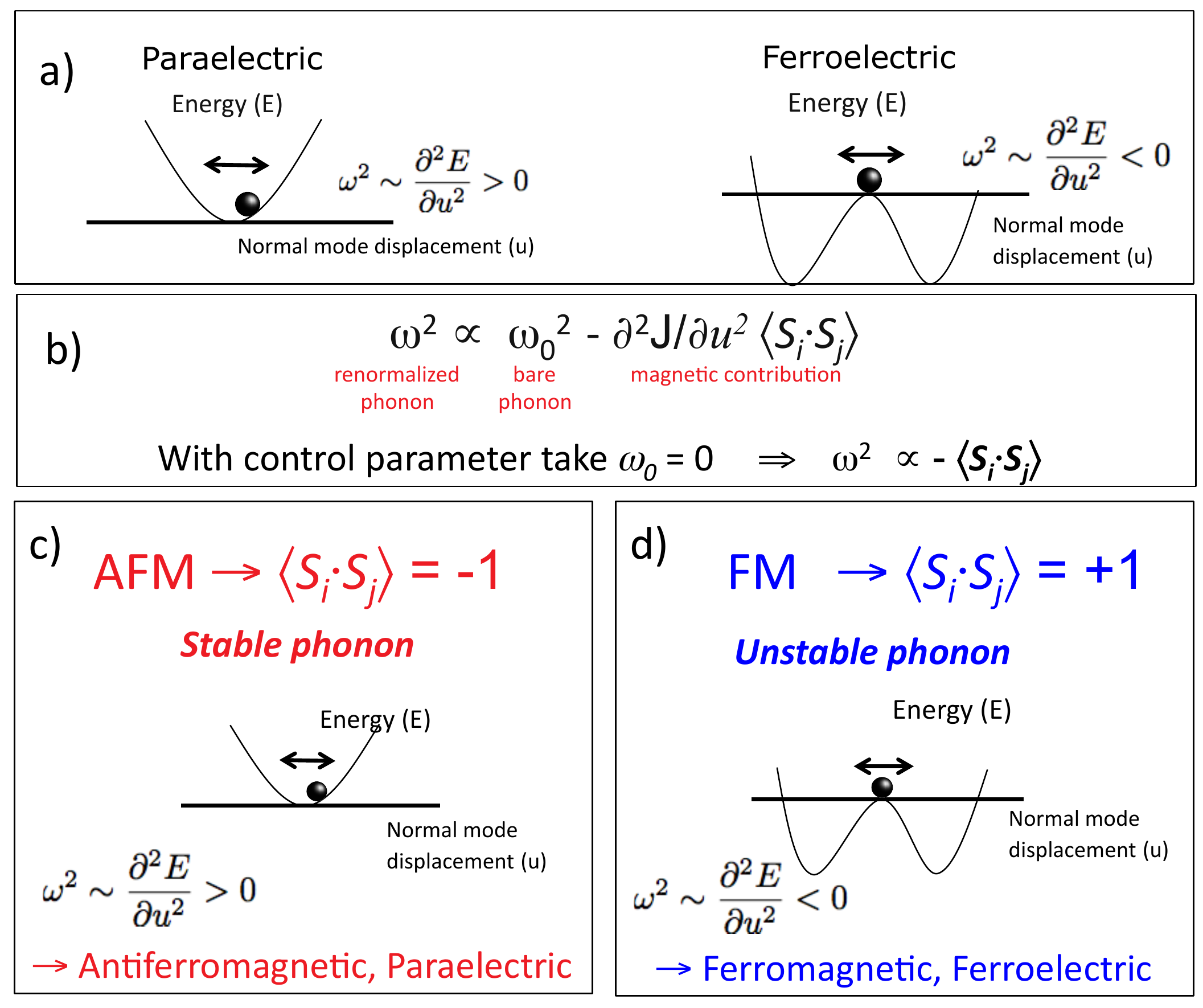} 
\end{center}
\caption{(a) Sketch of energy vs. normal mode displacement for stable (left) and unstable (right) phonon modes. (b) The phonon frequency squared can be separated into contributions from a bare phonon frequency (the frequency in the paraelectric state) and  a magnetic contribution that involves spin-spin correlations. (c) and (d) If the bare frequency is set to zero by an control parameter such as strain, the stability of the phonon mode depends on the sign of these correlations. }
\label{fig:phasecomp}
\end{figure}

In most materials the magnetic energy scale is much smaller than the structural one, therefore the frequency change of a phonon due to spin-phonon coupling is expected to be small. 
Recently, however, it was shown how this small spin-phonon coupling could
not only produce a large/colossal magnetocapacitive effect by mean of the
structural softness idea discussed in section \ref{lme} (the difference being,
\'{I}\~{n}iguez discussed a bilinear coupling while here it is a higher order
coupling) but also how it can be exploited to produce a large
magnetoelectric response.
 It was realized that if a system is close to a ferroelectric phase transition, the spin-phonon coupling can produce radically different behavior by driving the system back and forth across the structural/magnetic boundary. Elaborating further, we know that in the soft-mode theory of ferroelectricity the polar ground state is associated with an unstable polar phonon, Figure~\ref{fig:phasecomp}(a). The idea is to take a material in which this polar phonon is additionally coupled to the spin system and tune the bare phonon frequency to near zero, Figure~\ref{fig:phasecomp}(b). In this case the structural ground state depends on whether the dominant magnetic interactions are antiferromagnetic Figure~\ref{fig:phasecomp}(c), or ferromagnetic, Figure~\ref{fig:phasecomp}(d) (note the sign of the second derivative is system dependent). 

Hence, in the region where the bare phonon frequency is small, the
competition between the antiferromagnetic-paraelectric and the
ferromagnetic-ferroelectric states (or between the
antiferromagnetic-ferroelectric and the ferromagnetic-paraelectric states
if the sign of the second derivative is opposite) can lead to colossal
magnetoelectric response. Additionally, the boundary of this phase competition
region is bordered by a region of large/colossal magnetocapacitive
responses due to the softening of the lattice. The encouraging aspect of this design approach is that  the starting material is the typical magnetic insulator that is antiferromagnetic and paraelectric.

An example of a material with  spin-phonon coupling is EuTiO$_3$, which is a PE insulator that crystallizes in the perovskite structure in bulk. The Eu$^{2+}$ spins have G-type AFM order. Experiments on EuTiO$_3$ \cite{katsufuji01} showed that not only is there a suppression of the dielectric constant at the magnetic N\'eel  temperature, but also that an external magnetic field of sufficient strength to align the magnetic moments causes a sizable increase in the dielectric constant. First-principles calculations \cite{fennie06b,pentcheva07} verified that  spin-phonon coupling, as suggested by the experimentalists, is indeed the reason for this behavior.

In EuTiO$_3$, the magnetic ion Eu is on the A site of the perovskite structure, unlike typical magnetic perovskites such as SrMnO$_3$ or SrCoO$_3$. It is tempting to assume that the strength of exchange interactions does not depend on the ferroelectric mode, which mostly consists of B cation displacement. In order to explain the strong spin-phonon coupling that nevertheless exists in this material, Akamatsu et al.~\cite{akamatsu11}, employed first principles calculations and showed that there exists a hybridization between the half-filled Eu $f$ orbitals and the formally unoccupied Ti $d$ orbitals. This observation, combined with the fact that the bottom of the conduction band consists dominantly of Ti $d$ states, lead the authors to conclude that there exists an AFM superexchange interaction that is mediated via the Ti $d$ orbitals. The strength of this interaction depends strongly on the position of the Ti ion, and therefore the magnetic order is strongly coupled with the ferroelectric mode. This novel superexchange mechanism and the manner in which it couples with other lattice distortions, such as oxygen octahedron rotations, is a topic of active research, not only in the context of EuTiO$_3$ but also other rare earth - transition metal oxides \cite{akamatsu12,birol12}.

\begin{figure}[t]
\begin{center}
\includegraphics[width=0.45\textwidth]{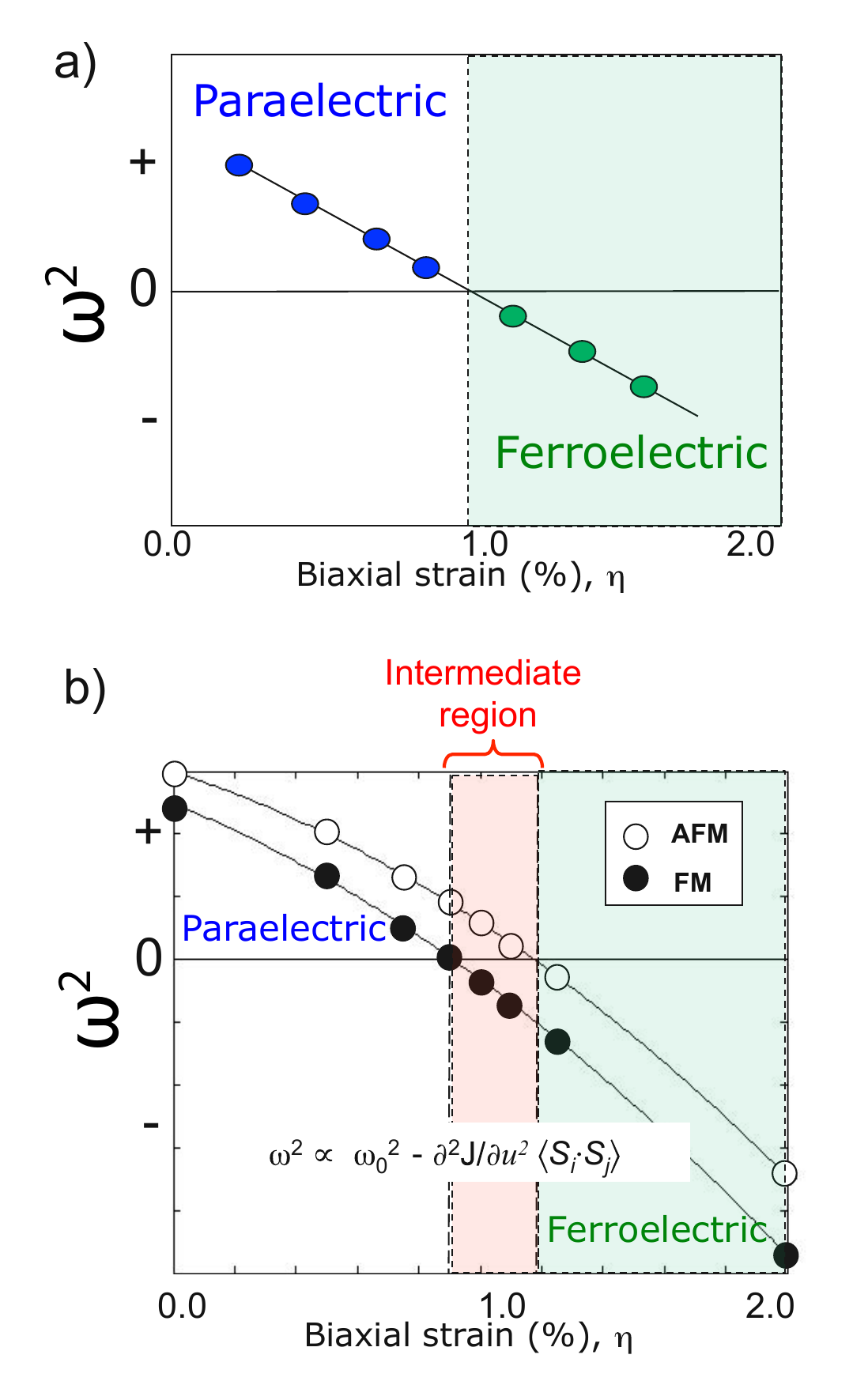} 
\end{center}
\caption{(a) Sketch of soft mode frequency squared vs. biaxial strain for a material that becomes ferroelectric under strain. (b) An equivalent sketch for a magnetic material that has spin - phonon coupling and a lower soft mode frequency in the FM state. There is a $\sim0.2\%$ wide strain range where the FM state has a ferroelectric instability but the AFM state does not.}
\label{fig:strain}
\end{figure}


So EuTiO$_3$ has the desired characteristics: its ground state is antiferromagnertic and paraelectric yet parallel spins soften the polar phonon. But how can the system be ``dialed'' into the region of the phase diagram where the phase competition takes place?
Epitaxial strain is widely used to stabilize new or meta-stable structural phases in transition metal oxides. In the case of perovskites it is well known that the polar soft mode is strongly sensitive to epitaxial strain and can even drive it from a paraelectric to a ferroelectric ground state, Figure~\ref{fig:strain}(a). A well known example is SrTiO$_3$, which is a quantum paraelectric in bulk but has been experimentally demonstrated to display ferroelectricity at  room temperature in thin film form under biaxial strain~\cite{haeni04}. 
What has recently been appreciated is that in a material with spin-lattice coupling, not only new structural phases but also new magnetic phases can be obtained by strain engineering.
%
Figure~\ref{fig:strain}(b) shows how the soft polar mode frequency varies with strain in an antiferromagnetic material that has spin - phonon coupling and for which the system with parallel spins, i.e., FM, has a softer mode frequency than in the AFM state, such as EuTiO$_3$. 

\begin{figure}
\begin{center}
\includegraphics[width=0.5\textwidth]{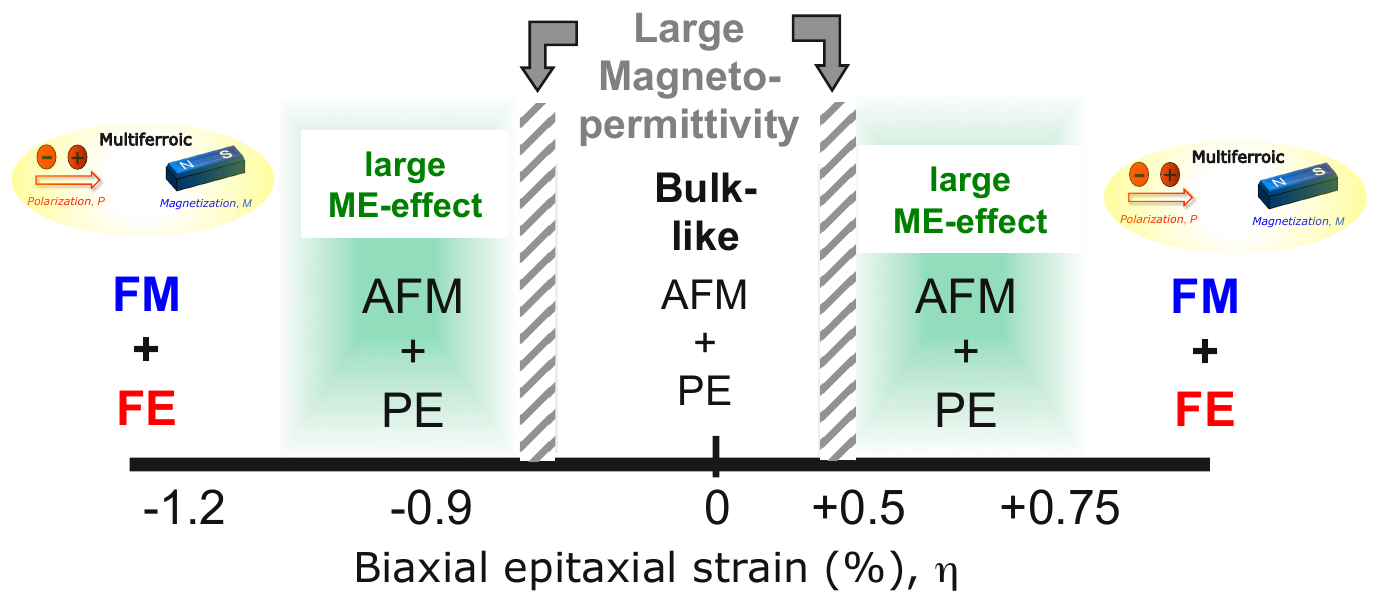} 
\end{center}
\caption{ Properties of EuTiO$_3$ under biaxial strain. (Horizantal axis is not drawn to scale.) Ground state is AFM - PE for small values of strain, and FM - FE for large values. There exist intermediate regions with large ME effect for both tensile and compressive values of strain.}
\label{fig:ETO_PD}
\end{figure}

We point out that there is a wealth of information in Figure~\ref{fig:strain}b that is only accessible from first-principles.
First, regardless of magnetic order, the soft mode is stable ($\omega^2>0$) for small
values of strain, and unstable ($\omega^2<0$) for large values. Next, as strain increases
and the lattice softens, there is a paraelectric region where the system
with parallel spins has a near-zero polar phonon frequency while this
phonon remains relatively hard in the ground state AFM. This is
a region of large/colossal magneto-capacitance. Notice, however, the exact
strain value at which the paraelectric to ferroelectric transition occurs
depends on magnetic order. 
Therefore there is an intermediate strain range for which a ferroelectric instability exists only if the magnetic state is ferromagnetic. 
It is in this intermediate region where there is the possibility for phase competition between and antiferromagnetic-paraelectric and ferromagnetic-ferroelectric phases. If the ground state remains AFM-PE, it should be possible to apply a magnetic field, which would align the spins, thereby driving the system into the ferroelectric phase (or apply an electric field to induce an electrical polarization thereby driving the system into the FM phase), obtaining a large magnetoelectric response.
The complete strain phase diagram of EuTiO$_3$, calculated from first principles, is presented in Fig. \ref{fig:ETO_PD}. Note that there exists an intermediate region with large magnetoelectric response for both compressive and tensile strains, and the ground state is ferromagnetic - ferroelectric for large strain whether it is compressive or tensile. The reason is that biaxial strain favors a ferroelectric state, and the ferroelectric displacement of the Ti atom suppresses the AFM exchange interaction. This prediction has been recently verified for EuTiO$_3$ thin films grown on DyScO$_3$ substrate, which provides a 1.1\% tensile strain~\cite{lee10}.

The subsequent experimental realization of a ferroelectric-ferromagnetic state in epitaxially-strained thin films of EuTiO$_3$ was a breakthrough in multiferroic design and a proof-of-principle material~\cite{lee10} (see also the article in this issue by Martin and Schlom).  Several fundamental scientific challenges remain, however, before materials useful for practical devices can be identified.  First, as discussed the present method relies on the ability to apply large epitaxial strains to a material.  Such epitaxial strain engineering is readily possible today where the application of 1-2$\%$ biaxial tensile and compressive strains is routine.  This method, however, requires growing a thin film coherently on a substrate of a larger/smaller lattice constant.  The maximum thickness for which the film can be grown coherently with the substrate is determined as a balance between the volume strain energy of the film and the energetic cost of introducing misfit dislocations or cracks.  For thicknesses on the order of 50 nm for oxide films strained by 1$\%$, misfit dislocations are typically introduced, relaxing the strain imposed across the interface.  For a variety of technological reasons, it would be beneficial to identify a control mechanism (a ``knob'' ) that can feasibly work for larger volumes of material.  A second challenge concerns the temperature behavior of the material.  Ideally, for practical applications in electronic, magnetic, or optical devices, the characteristic operating temperature of the material should approach room temperature.  In principle the novel mechanism of which EuTiO$_3$ is a realization could operate at room temperature, however EuTiO$_3$ itself orders magnetically at 5.5 K.
 It is therefore important to identify alternative materials with higher magnetic ordering temperatures (above room temperature, preferably). As we have been discussing in this review, first-principles methods are playing an important role in the search for such systems.

\subsection{Raising the temperature: $f\rightarrow d$}

Recently, Lee and Rabe used first-principles calculations to study the spin-phonon coupling in a series of perovskites SrMO$_3$ (M = V, Cr, Mn, Fe, Co)~\cite{lee11}. In these materials, the magnetism is due to 3$d$ electrons, which hybridize much more strongly with their environment than 4$f$ electrons. Exchange interactions in SrMO$_3$ are thus stronger than in the rare earth titanates (such as EuTiO$_3$), leading to higher magnetic ordering temperatures. In particular, for SrMnO$_3$ and SrCoO$_3$, the transition temperatures are $233$ K and $305$ K, respectively.  The calculations of Lee and Rabe indicated large spin-phonon coupling for the soft polar mode of SrMnO$_3$ and SrCoO$_3$ crystals \cite{lee11}. Furthermore, for SrMnO$_3$ the polar mode became unstable on changing the magnetic configuration from AFM to FM, indicating a FE transtion. In the case of SrCoO$_3$, the system is metallic and the magnetic ground state is ferromagnetic. Spin-phonon coupling destabilizes a zone center mode (which would be polar if the system was not a metal) when G-type AFM ordering is imposed.

\begin{figure}[t]
\begin{center}
\includegraphics[width=0.45\textwidth]{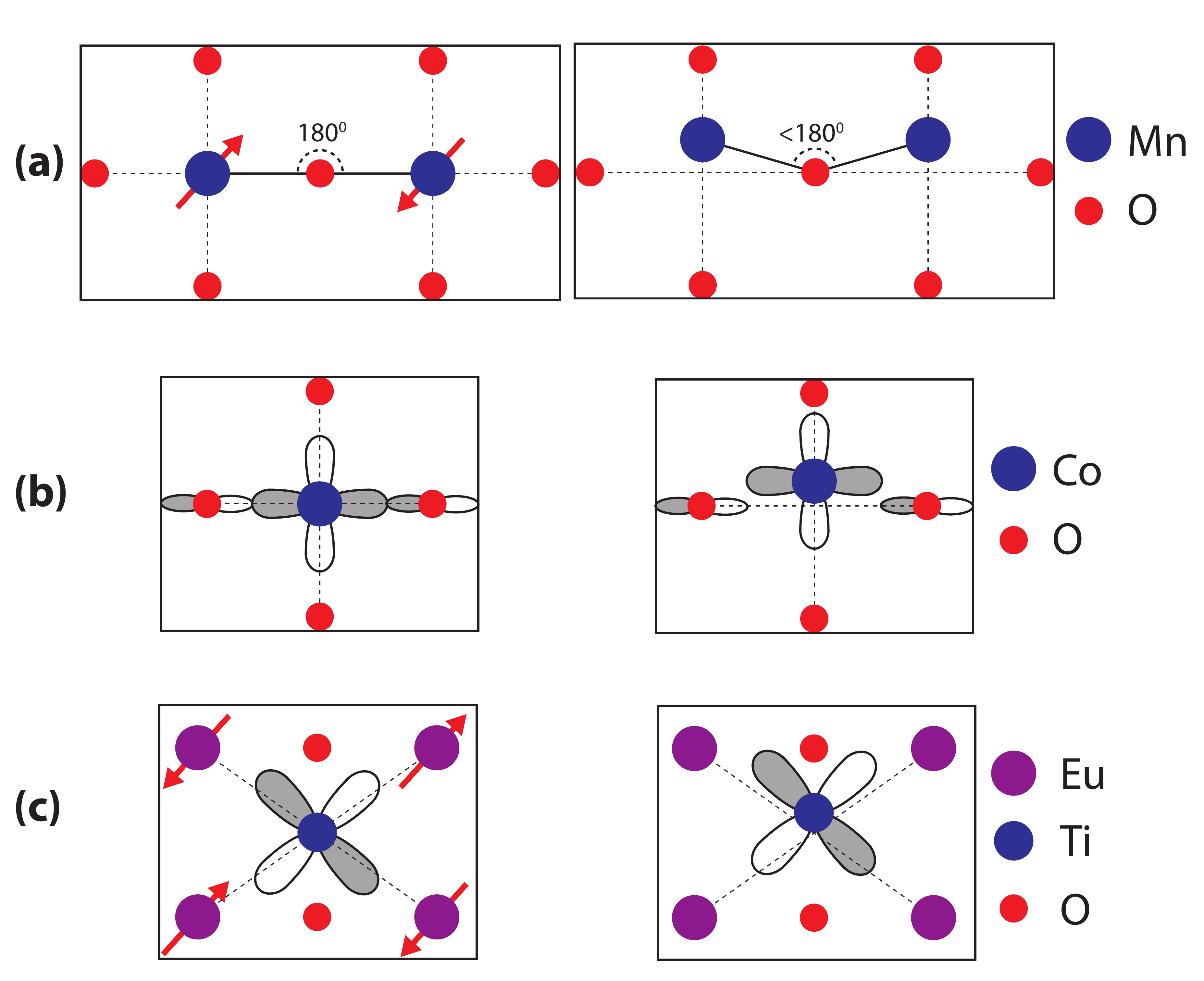}
\end{center}
\caption{a) Schematic picture illustrating the origin of the spin-phonon coupling for the polar mode in SrMnO$_3$. In the PE cubic state, the Mn-O-Mn angle is 180$^\circ$ and this maximizes the AFM superexchange interaction. The FE distortion reduces the the Mn-O-Mn angle and hence reduces the strength of the exchange interaction. b) As in a) but for SrCoO$_3$. The overlap between Co-$d$ and O-$p$ orbitals determines the strength of exchange interaction. The FE distortion decreases this overlap and hence the strength of the interaction. c) Illustration of the spin-phonon coupling in EuTiO$_3$. The Ti-mediated superexchange between the Eu $4f$ spins depends on the fourth power of the hopping amplitude $t_{fd}$ to the Ti $d$ orbitals that are (partly) directed towards Eu ions. Displacement of the Ti ion from the center of the unit cell due to the polar phonon mode changes $t_{fd}$ and hence weakens the exchange interaction.}
\label{spc-fig2}
\end{figure}

Lee and Rabe also identified the microscopic mechanisms responsible for the the spin-phonon coupling in SrMnO$_3$ and SrCoO$_3$. For both materials, the mechanism is directly related to the exchange interaction between spins. For SrMnO$_3$, the ground state is G-type AFM and the Mn atoms interact through superexchange, which is mediated by the oxygen atoms. The superexchange mechanism leads to the AFM nearest-neighbor exchange parameter given by,
\begin{equation}
J_{nn} \sim \frac{t^2}{\Delta-J_H}.
\label{spc-eq5}
\end{equation}
$J_H$ is the Hund coupling, $\Delta$ is the energy difference between the oxygen 2\emph{p} and Mn 3\emph{d} orbitals and $t$ is the hopping integral between Mn nearest neighbors, which is proportional to the effective wave function overlap between the two Mn atoms through their shared oxygen. The hopping integral, and therefore the magnitude of $J_{nn}$, is maximized for the cubic PE structure where the Mn-O-Mn angle is 180$^\circ$ (see Fig. \ref{spc-fig2}(a), left). In the FE structure, Mn atoms move toward one of the neighboring oxygens (see Fig. \ref{spc-fig2}(a), right) thereby decreasing the Mn-O-Mn angle, which leads to a decrease of $J_{nn}$. In addition, the reduction of the bond length between the Mn atom and one of the oxygens enhances the hybridization between the Mn 3\emph{d} and O 2\emph{p} orbitals, which increases $\Delta$ and decreases $J_{nn}$ further. This reduction of the AFM exchange interaction in the FE state is responsible for the spin-phonon coupling. Indeed, if the FM state is enforced then the energy cost for the FE distortion decreases because it diminishes the energetically unfavorable exchange interaction. 

For metallic SrCoO$_3$, the FM ground state ordering was shown to be a result of a mechanism that is similar to the double exchange interaction \cite{potze95, zener51}. Similarly, as in the case of AFM superexchange for SrMnO$_3$, the Zener double exchange interaction, which depends sensitively on $d$-$p$ hopping, is maximized when the Co-O-Co angle is 180$^\circ$ and therefore it is diminished when the Co ion is displaced due to a phonon [see Fig. \ref{spc-fig2}(b)]. This leads to the softening of this phonon mode as the G-type AFM ordering is enforced \cite{lee11}.

\begin{figure}[ht]
\begin{center}
\includegraphics[width=0.45\textwidth]{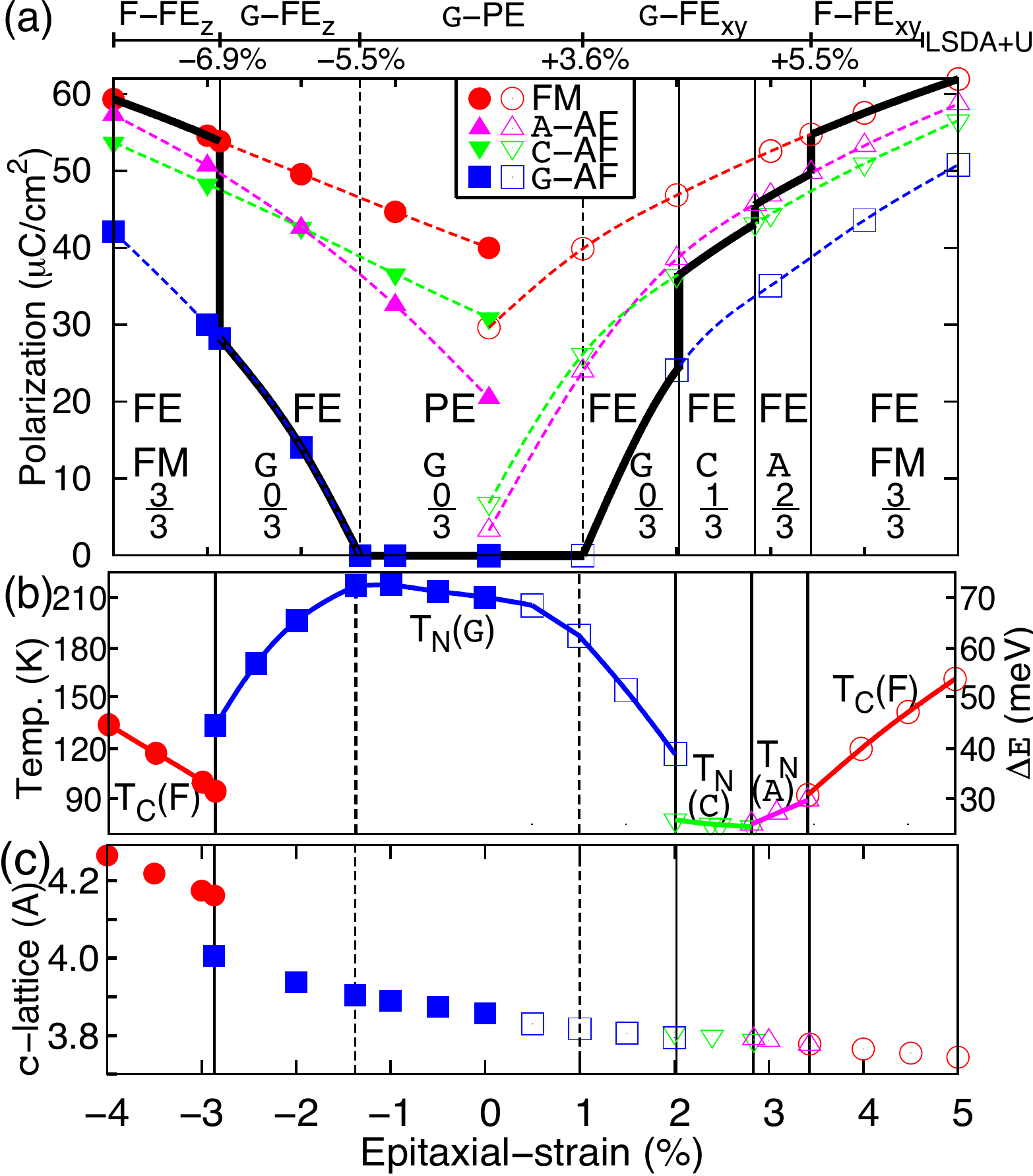} 
\end{center}
\caption{ Properties of SrMnO$_3$ under biaxial strain. (Reproduced from Ref. \cite{lee10b}.) (a) Polarization of various phases, and the ground state (black). (b) Magnetic ordering temperature of the ground state. (c) c lattice constant. Copyright (2010) by the American Physical Society.}
\label{fig:srmno3}
\end{figure}

Lee and Rabe also showed that both SrCoO$_3$ and SrMnO$_3$ undergo various phase transitions as a function of strain \cite{lee10b, lee11c}. SrCoO$_3$, a ferromagnetic metal, becomes an antiferromagnetic insulator under large enough tensile or compressive strain. SrMnO$_3$, which is an antiferromagnetic paraelectric, becomes a ferromagnetic ferroelectric under large enough tensile or compressive strain. In addition to these phases, SrMnO$_3$ can be stabilized with various different antiferromagnetic states and has a magnetic ordering temperature of $\sim 100$ K (See Fig. \ref{fig:srmno3}).

The transition of SrMnO$_3$ to a FE - FM state can be explained in the same way as EuTiO$_3$: strain favors ferroelectricity, which suppresses the AFM exchange interactions. Experimental observation of a similar spin - lattice coupling in perovskite Sr$_{1-x}$Ba$_x$MnO$_3$ \cite{sakai11}  supports this explanation. Increasing the Ba concentration in Sr$_{1-x}$Ba$_x$MnO$_3$ increases the lattice constant and drives the system ferroelectric for $x\gtrsim0.45$. The critical temperature for ferroelectricity is larger than that of antiferromagnetism and the polarization decreases significantly upon magnetic ordering. The authors explain this reduction in polarization by the fact that the energy gain due to AFM order is largest when the Mn - O - Mn angle is 180$^\circ$, so the magnetic ordering favors a nonpolar state.

\subsection{Spin-phonon coupling from first principles: which exchange-correlation functional to choose?}
First-principles calculations of magnetism in transition metal oxides are usually performed using the DFT$+U$~\cite{anisimov97} method, where a Hubbard-type $U$ term (partially) takes into account strong correlations between localized electrons. Though practical, the approach does have some serious drawbacks, including the fact that there is no obvious way to choose an appropriate value for $U$ (there has been some progress on approaches in which $U$ is evaluated self-consistently for the system of interest~\cite{kulik06}). The standard procedure usually involves varying $U$ until some chosen quantity (such as the band gap, for example) matches the experimental value. Care must be taken to check the $U$ dependence of the calculated properties of interest, as we will discuss further below. There are (beyond DFT) techniques which explicitly capture the interaction between localized electrons, such as Dynamical Mean Field Theory (DMFT) \cite{georges96,kotliar06} or Quantum Monte Carlo (QMC)~\cite{foulkes01}. DMFT in particular has provided many insights into the electronic structure of strongly correlated materials~\cite{pavarini04,ramakrishnan04,biermann05,han11}. However, as powerful as these techniques are, the calculation of forces is difficult, an obvious problem if the aim is to study spin-phonon coupling.  

Hong and co-workers~\cite{hong12} studied the spin-phonon coupling in a series of perovskite transition metal oxides using both PBEsol+$U$ and the Heyd-Scuseria-Ernzerhof (HSE) hybrid functional~\cite{heyd06}. Hybrid functionals incorporate a fraction of exact exchange from Hartree-Fock theory and are generally accepted to provide an improved description of the electronic structure of many materials; the Heyd-Scuseria-Ernzerhof (HSE) hybrid functional~\cite{heyd06,henderson11} in particular has been shown to provide an improved description of the properties of transition metal oxides~\cite{archer11} (including phonon frequencies~\cite{wahl08,franchini10}) with the advantage of being less computationally demanding than other hybrids, such as PBE0. However, the calculation of some quantities, such as phonon frequencies can still be prohibitively expensive. Hong, \textit{et al.,} developed a procedure in which the value of $U$ is determined by fitting to the HSE results for the difference in total energy between different magnetic configurations. In testing their approach, they found that the magnitude of spin-phonon coupling was very sensitive to the choice of $U$ for certain modes. They showed that the frequency shifts between different magnetic arrangements calculated with PBEsol+$U$ (with the $U$ fitted to the HSE data) were in very good agreement with the frequency shifts calculated with HSE directly. This procedure therefore represents a practical and convenient way to increase the reliability of such calculations without increasing the computational expense.

\section{The magnetoelectric effect in multiferroic heterostructures and superlattices}
\label{superlattice}

The discussion so far has been focused on homogeneous materials and intrinsic mechanisms. Recent advances in epitaxial growth techniques, however, have opened paths to create artificial multiferroic systems by forming layered heterostructures composed of magnetic and dielectric materials. Such composite multiferroics have attracted much attention, since they may exhibit a magnetoelectric (ME) effect that is significantly larger than in single-phase multiferroics. In general, we can distinguish between two types of ME coupling that occur in such heterostructures: (1) strain-mediated ME effect, and (2) interfacial ME effect. In the former type, the only effect of the interface is to strain the heterostructure constituents, which allows for control of their bulk properties. In this context, bilayers composed of piezoelectric and piezomagnetic materials have been intensively investigated. Here, the piezoelectric (piezomagnetic) layer can be strained by  application of an electric (magnetic) field. This strain is `transmitted' to the neighboring piezomagnetic (piezoelectric) layer and changes its magnetization (polarization) leading to ME coupling \cite{suchtelen72,mitoseriuab06,stefanita08}. First-principles studies of this type of ME effect are usually concerned with the strain dependence of bulk properties of piezoelectric \cite{janolin12} and piezomagnetic \cite{lukashev08} materials and will not be discussed here. Instead, we focus on the second type of ME coupling, which is purely due to interfacial effects.

Using first-principles calculations it was shown that a large ME effect is produced at the Fe/BaTiO$_3$ interface\cite{duan06,fechner0878}. The TiO$_2$-terminated interface is stable\cite{fechner0877} and a magnetic moment opposite in direction to the Fe magnetization is induced at the interfacial Ti atoms. The value of the Ti moment depends, however, on the polarization of BaTiO$_3$ (BTO) being equal to 0.18 $\mu_B$ for the polarization pointing away from the interface and 0.40 $\mu_B$ for the polarization pointing toward the interface \cite{duan06}. Hence, application of an electric field that switches the polarization of the BTO leads to a change of interfacial magnetization. In order to quantify this effect, a surface ME coefficient was introduced, $\alpha_S = \Delta M_s/E_c$, where $\Delta M_s$ is the change of interface magnetization and $E_c$ is the coercive field of the ferroelectric layer. Using the above results and the experimental value for the the coercive field of BTO, the surface ME coefficient was estimated  to be 2 $\times$ 10$^{-10}$ G cm$^2$/V \cite{duan06}. This value, however, strongly depends on the oxygen stoichiometry at the interface \cite{fechner09}.

\begin{figure}
\begin{center}
\includegraphics[scale=0.5]{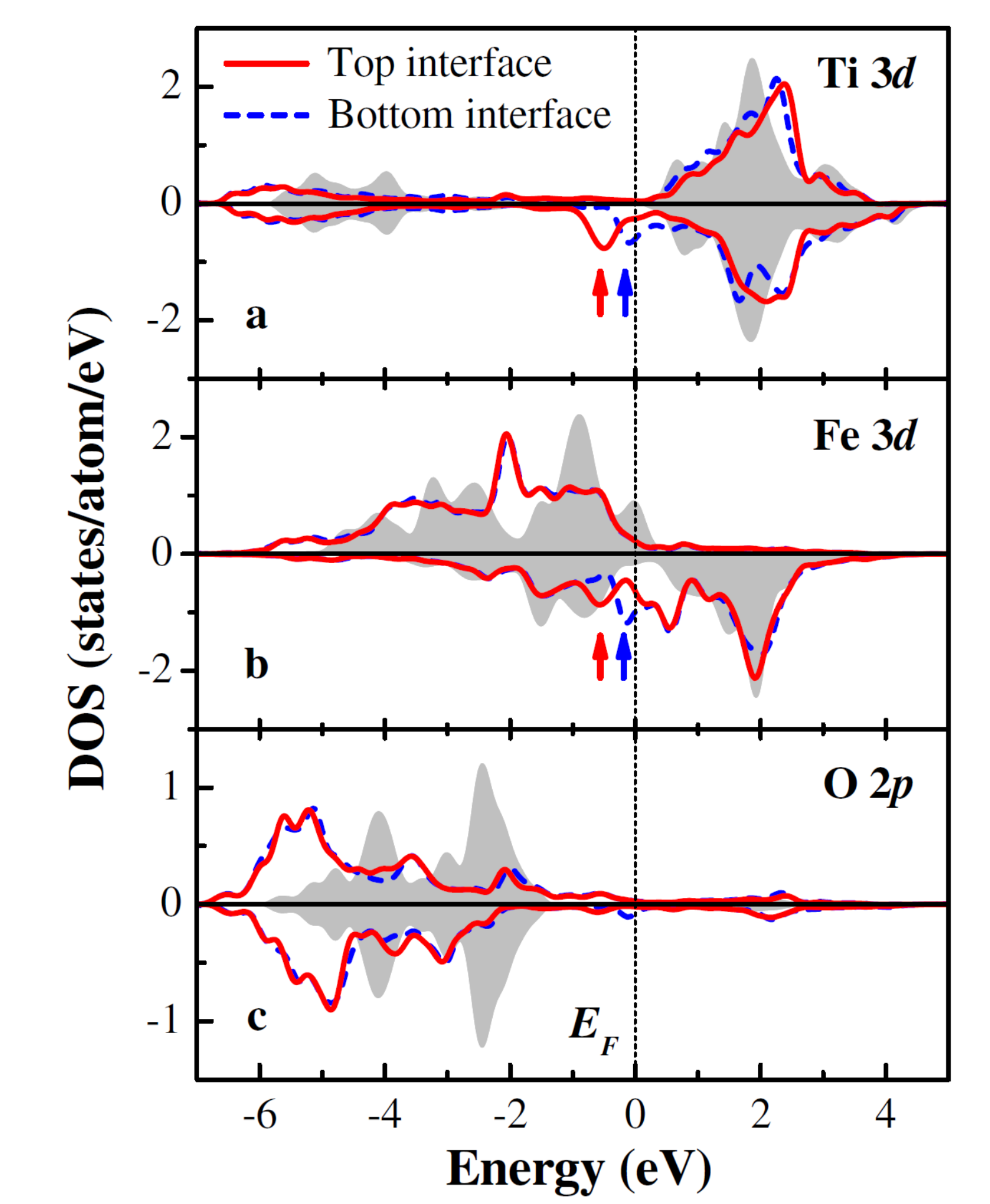} 
\end{center}
\caption{Orbital-resolved DOS for interfacial atoms in a Fe/BaTiO$_3$ multilayer for net polarization of BaTiO$_3$ layer pointing toward (away from) the top (bottom) interface: (a) Ti 3\textit{d}, (b) Fe 3\textit{d}, and (c) O 2\textit{p}. Majority- and minority-spin DOS are shown in the upper and lower panels, respectively. The solid and dashed curves correspond to the DOS of atoms at the top and bottom interfaces, respectively. The shaded plots are the DOS of atoms in the central monolayer of (b) Fe or (a),(c) TiO$_2$ which can be regarded as bulk. The vertical line indicates the Fermi energy (E$_F$). From Ref.\cite{duan06}. Copyright (2006) by the American Physical Society.}
\label{cm-fig1}
\end{figure}

The physical mechanism responsible for this interface ME effect can be understood by analyzing the orbital-resolved local density of states (DOS)\cite{duan06}. Fig. \ref{cm-fig1} shows that the Ti 3$d$ band is centered at about 2 eV above the Fermi level and overlaps strongly with the minority Fe 3$d$ band, which has a significant weight in this energy region. This leads to hybridization between the Ti 3$d$ and and minority Fe 3$d$ states, which results in formation of bonding states that are pushed down in energy and peaked just below the Fermi level (the peaks are indicated by arrows in Fig. \ref{cm-fig1}(a) and \ref{cm-fig1}(b)). This results in a larger occupation of Ti minority states as compared to the majority states (the majority and minority channels are determined by Fe), resulting in a magnetic moment on Ti that is antiparallel to the magnetic moment of Fe \cite{duan06}. If the polarization in BTO points toward the interface the Fe-Ti bond length becomes shorter, which enhances the hybridization and thus pushes the minority bonding state to even lower energies so that it becomes more populated, resulting in a large magnetic moment on Ti. On the other hand, if the polarization points away from the interface, the Fe-Ti bond length increases, which weakens the hybridization and pushes the minority bonding state to higher energies so that it is less populated and the Ti magnetic moment decreases \cite{duan06}. First-principles calculations have revealed that such a `hybridization-driven' ME effect also occurs for many other heterostructures, including Fe/PbTiO$_3$ (PTO) \cite{fechner0878}, Co$_2$MnSi/BaTiO$_3$ \cite{yamauchi07} and Fe$_3$O$_4$/BTO \cite{niranjan08}.

Rondinelli \textit{et al.} \cite{rondinelli08} proposed another mechanism to drive an interfacial ME effect based on the response of free electrons in a metal to an external electric field. When a metal is placed into an electric field ($E$), the free electrons flow towards the surface and screen the electric field over the screening length of the metal. In the case of a ferromagnetic metal this accumulated surface charge is spin polarized, which induces a net change in magnetization at the interface. Such an ME effect driven by spin-dependent screening was demonstrated to occur at surfaces of ferromagnetic metals including Fe, Co, Ni\cite{duan08}, SrRuO$_3$\cite{niranjan09}), and halfmetallic (Fe$_3$O$_4$) \cite{duan09}, but the estimated ME coefficient is very small. The ME effect may, however, be substantially enhanced at interfaces between a metallic ferromagnet and an insulator due to the dielectric response of the insulator that effectively increases the electric field that must be screened by the ferromagnet. In particular, the induced surface charge, $\sigma = \varepsilon_0\varepsilon E$, is proportional to the dielectric constant ($\varepsilon$) of the insulator. Therefore, a large ME effect can be expected at the interfaces between metallic ferromagnets and high-$\varepsilon$ dielectrics. Indeed, first principles calculations for the SrRuO$_3$/SrTiO$_3$ (SRO/STO) interface in the presence of an electric field revealed a large ME response with $\alpha_S \approx$ 2 $\times$ 10$^{-12}$ G cm$^2$/V \cite{rondinelli08}. The origin of this effect was shown to be the spin-polarized screening which changes the occupation of the interfacial Ru-4\textit{d} states, resulting in a change in local magnetic moment. \cite{rondinelli08}

An even stronger ME effect can be expected at the interface between ferromagnetic metals and ferroelectrics. In this case, changing the direction of polarization by application of an electric field changes the sign of polarization charges at the interface and therefore modifies the induced surface charge in the ferromagnet at the interface. As the induced surface charge is spin-polarized, this induces a change in the interface magnetization. First-principles calculations show that this mechanism leads to the ME effect at the SrRuO$_3$/BaTiO$_3$ interface with $\alpha_S \approx$ 2 $\times$ 10$^{-12}$ G cm$^2$/V 2.3 $\times$ 10$^{-10}$ G cm$^2$/V \cite{niranjan09} , which is two orders of magnitude larger than for SrRuO$_3$/SrTiO$_3$. 
 
As discussed above, the screening of an electric field at metallic surfaces changes the carrier concentration in the surface region and thus can be thought of as electrostatic doping. In particular, if we consider the surface of a metallic alloy then one can effectively change the position of the system on the concentration phase diagram by applying an external field. This may lead to a dramatic effect if the system sits close to the phase boundary, since then it would be possible to switch between two phases using an electric field. In particular, a gigantic ME effect can be achieved when the system sits at the boundary between FM and AFM phases.

\begin{figure}[h]
\begin{center}
\includegraphics[width=0.9\hsize]{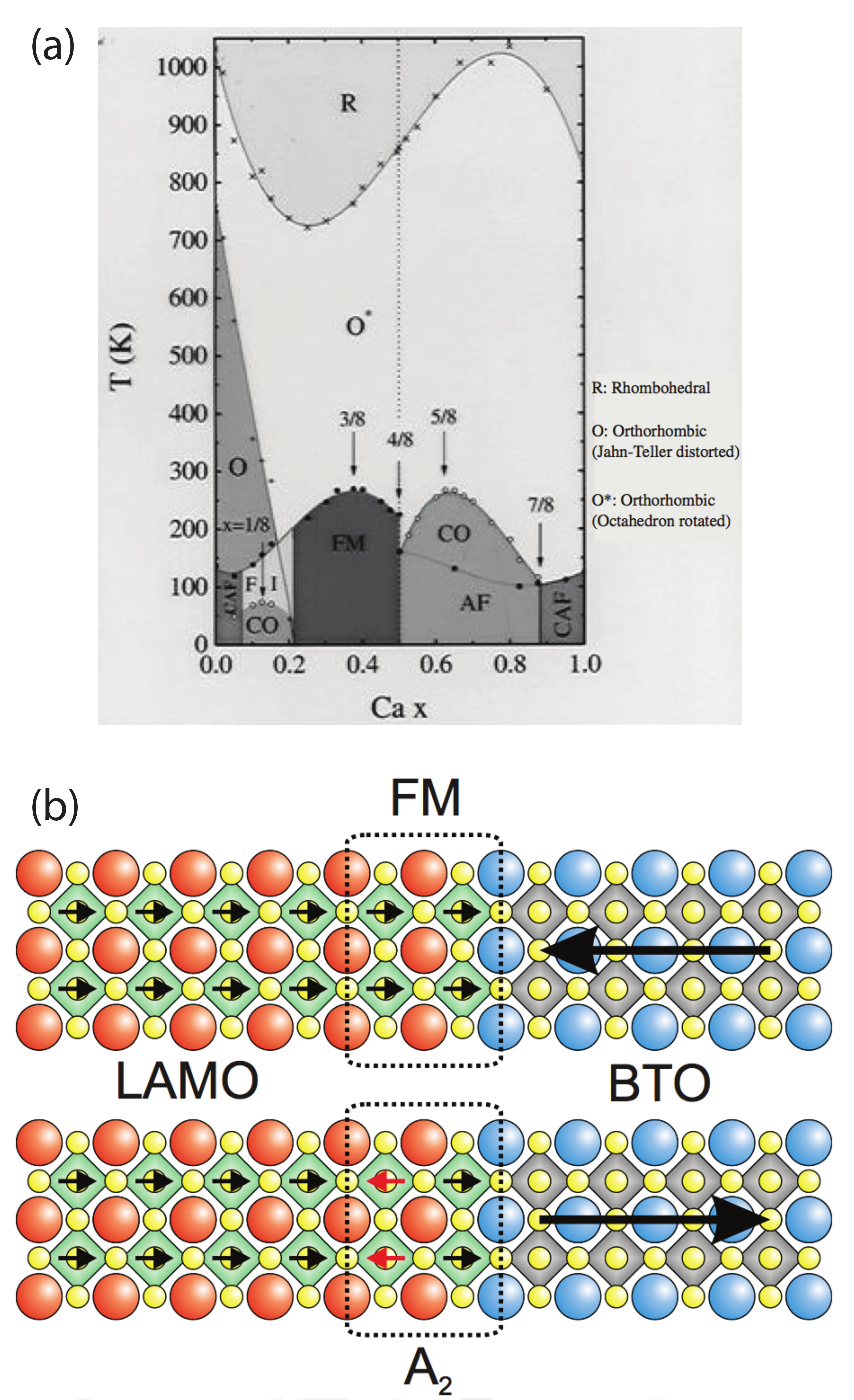} 
\end{center}
\caption{(a) Phase diagram of La$_{1-x}$Ca$_x$MnO$_3$ as a function of doping concentration $x$ and temperature T. From Ref.\cite{ramakrishnan07}. (b) Electrically induced magnetic reconstruction at the LAMO/BTO interface; the predicted change in order of the Mn magnetic moments (small arrows) from FM to A-type antiferromagnetic (A2) as the ferroelectric polarization (large arrows) of the BTO is reversed. From Ref.\cite{burton09}. Copyright (2009) by the American Physical Society.}
\label{LSMO}
\end{figure}

La$_{1-x}$A$_x$MnO$_3$ (LAMO) (A = Ca, Sr or Ba) alloys show a rich phase diagram as a function of doped carrier concentration, which consists of different resistive, magnetic as well as structural phases. As an example we show a phase diagram for La$_{1-x}$Ca$_x$MnO$_3$ (see Fig.~\ref{LSMO}(a)). For our purposes, the most interesting is the phase boundary between the FM metallic phase and the (AFM) insulating phase at concentrations of about 0.5. Based on the above discussion, an external electric field may switch between these phases at the surface of La$_{0.5}$Ca$_{0.5}$MnO$_3$. Similarly, as in the case of induced surface magnetization, a larger effect may be expected at the interface with an FE material since an electric field produced by polarization charges at the surface of a ferroelectric can be much larger than experimentally accessible electric fields.
Indeed, first principles calculations for (LAMO)/BTO interfaces have shown that switching the polarization of BTO by an electric field switches the magnetic order at the alloy side of the interface between AFM and FM (see Fig.~\ref{LSMO}(b)) \cite{burton09}. 

Heterostructures built from metallic and ferroelectric materials
provide a promising avenue of research, however one must approach the
first-principles treatment of such systems with extra care. The
well-known `band gap error' (the underestimation of the band gaps of
semiconductors and insulators \cite{perdew83,
sham83}) may in particular complicate
calculations of interfaces. Stengel and co-workers have discussed in
detail \cite{stengel11} that the possible pathological alignment of the Fermi level of
the metal with the conduction band of the ferroelectric at
an interface could result in the incorrect prediction of a metallic heterostructure. The ferroelectric layer, which should be insulating, becomes metallic because its conduction band is occupied by free charges from the metal. Fundamental studies of the physics of such interfaces, as
well as studies of the accuracy of different theoretical treatments
(the accuracy of hybrid functionals for band alignment, for example),
are currently active areas of research.


Recently, manipulation of magnetocrystalline anisotropy (MCA) by applying an electric field has been achieved for many FM metallic interfaces (surfaces) \cite{weisheit07, maruyama09, shiota09} as well as at the interface (surface) of dilute magnetic semiconductors \cite{chiba08,ohta09}. Interestingly, a significant change in perpendicular MCA energy ($\sim$ 40 \%) by an applied voltage at  the Fe/MgO interface \cite{maruyama09} and moreover the direct switching from out-of-plane to in-plane magnetization has been achieved in Au/Fe$_{0.8}$Co$_{0.2}$/MgO \cite{shiota09}. Further, a modification of the out-of-plane magnetic anisotropy via piezoelectric response has been shown for the epitaxially grown magnetite films on the BaTiO$_3$ substrate \cite{vaz09}. 

\begin{figure}
\begin{center}
\includegraphics[scale=0.24]{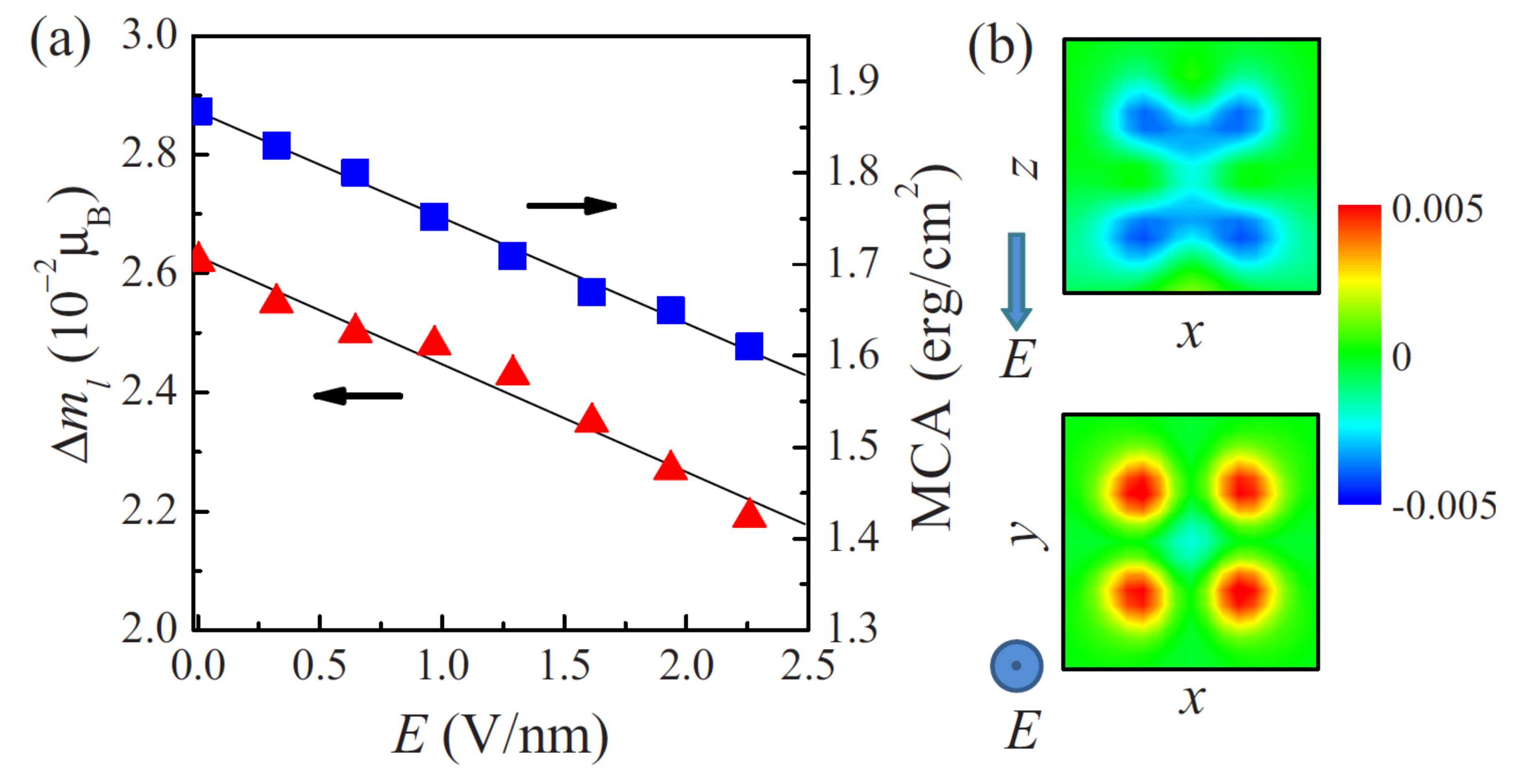} 
\end{center}
\caption{(a) Magnetic anisotropy energy (MCA, squares) and orbital moment anisotropy ($\Delta m_l$, triangles) of the Fe/MgO(001) interface as a function of electric field in MgO. (b) Induced charge density $\Delta \rho = \rho (E) - \rho (0)$, in units of $e/\AA^3$, at the interfacial Fe atom for E=1.0 V/nm in the x-z (010) plane (top panel) and the x-y (001) plane (bottom panel). Reprinted with permission from Ref. \cite{niranjan10}. Copyright (2010) American Institute of Physics.}
\label{MCA}
\end{figure}

Modulation of MCA by an electric field was also reported by \textit{ab initio} studies \cite{duan08, duan08apl,nakamura09,tsujikawa09,zhang09,niranjan10}. In Fig.~\ref{MCA}(a) we show a calculated electric field dependence of MCA for the Fe/MgO interface \cite{niranjan10}. In order to understand this effect, we remind the reader again that MCA is a result of  spin-orbit coupling. Using second-order perturbation theory and neglecting the spin-flip terms, in the case of a fully occupied majority $d$ spin channel (as in case of Fe), the MCA energy can be written as,
\begin{equation}
E^{MCA} = E[100] - E[001] = -\dfrac{\xi}{4}(S_x\langle L\rangle_x - S_z\langle L\rangle_z),
\label{mca_energy}
\end{equation}
where $\xi$ quantifies the strength of the coupling between $\vec{L}$ and $\vec{S}$. Equation \ref{mca_energy} shows that the MCA energy varies with the orbital moment anisotropy ($\Delta m_l$). Indeed, a change of the MCA for the Fe/MgO interface is accompanied by a change of the orbital moment anisotropy (see Fig.~\ref{MCA}(a)) \cite{niranjan10}. The origin of the change in the orbital moment anisotropy is found to be the relative change in the occupation of the \textit{d} orbitals. As shown in Fig.~\ref{MCA}(b), when the applied electric field is away from the Fe layer, electrons want to move away from the interface region and therefore the occupation of the $d_{xz}$ and $d_{yz}$ orbitals, which have a large charge density at the interface, is reduced. At the same time, the occupation of the $d_{xy}$ orbitals (which have a smaller charge density at the interface) increases \cite{niranjan10}. This leads to a decrease in $\langle L\rangle_z$, which results in a reduction of $\Delta m_l$ and hence $E^{MCA}$ (see Fig.~\ref{MCA}(a)). A similar mechanism also explains the change in MCA at FM metal surfaces \cite{duan08,nakamura09}. On the other hand, for the Fe/BaTiO$_3$ interface it was shown that the change of the orbital moment anisotropy (and therefore MCA) is due to modification of the hybridization between Fe-3\textit{d} states and Ti-3\textit{d} states induced by the change of Fe-Ti bond length under polarization reversal \cite{duan08apl} .

\section{Outlook} 
As the field moves forward, the focus of first-principles researchers working on the functional properties of complex oxides is increasingly on the properties of strongly correlated electron materials. 
Various novel phenomena are observed in strongly correlated oxides, high temperature superconductivity and colossal magnetoresistance being just two examples. Additional degrees of freedom created by orbital ordering lead to very rich phase diagrams involving interplay of both lattice and magnetism with electronic structure and additional complexity \cite{dagotto05}.  
We briefly touched on the technical difficulties associated with the first-principles treatment of such materials and mentioned a couple of the advanced methods currently available for calculations on strongly correlated electron systems. The problem is that these advanced techniques are currently the province of a relatively small group of experts and therefore need to become much more widely available and adopted if we are to bring the full force of the first-principles armory to bear on the study of magnetoelectric phenomena in strongly correlated materials. Luckily, the resurgence of interest in the nickelates and cuprates has pushed these issues to the forefront of first-principles research on complex oxides.\cite{dagotto05, dagotto08, boris11, jang11, mannhart10}

We thank Jorge \'{I}\~{n}iguez and Sinisa Coh for there valuable comments. T.B., N.A.B, S.G. and C.J.F. were supported by DOE-BES under Award Number DE-SCOO02334.  E.H.S was supported by DOE-BES under Award Number DE-SC0005032. A.T.M$.$ was supported by NSERC of Canada and by the NSF (No. DMR-1056441). H.D. was supported by Penn State NSF-MRSEC grant number DMR 0820404.  A.L.W. was supported by the Cornell Center for Materials Research with funding from the NSF MRSEC program, cooperative agreement DMR 1120296.

\section{Appendix}
\subsection{\label{section:DM}Appendix I: The Dzyaloshinskii-Moriya (DM) interaction}
The focus of many studies on magnetoelectric materials has been the the Dzyaloshinskii-Moriya (DM) interaction, also known as antisymmetric exchange (as opposed to symmetric Heisenberg exchange). The DM interaction between two spins $\vec{S}_i$ and $\vec{S}_j$ has the form 
\begin{equation}
E_{ij}^{DM}=\vec{d}_{ij}\cdot\left(\vec{S}_i\times\vec{S}_j\right)
\label{equ:DMmic}
\end{equation}
where $\vec{d}_{ij}$ is the DM vector. In 1957, Dzyaloshinskii pointed out \cite{dzyaloshinskii57} that a free energy contribution of the form 
\begin{equation}
F^{DM}=\vec{D}\cdot\left(\vec{M}\times\vec{L}\right),
\label{equ:DMmac}
\end{equation}
where $\vec{M}$ is the magnetization, $\vec{L}$ is the antiferromagnetic vector and $\vec{D}$ is a material specific vector, is allowed by symmetry in certain materials. Shortly after this, Moriya \cite{moriya60} showed that the spin-orbit correction to Anderson superexchange leads to the microscopic interaction in Equation \ref{equ:DMmic}, which in turn leads to the free energy contribution in Equation \ref{equ:DMmac}. 

Since the DM interaction is a relativistic effect, it is usually weak in the sense that the energy scale it involves is much smaller than the energy scale of Heisenberg exchange. However, it is still important because it gives rise to the weak ferromagnetism (wFM) observed in some antiferromagnets: any nonzero $\vec{D}$ causes the spins in an antiferromagnet to cant (typically by only a few degress) in order to minimize $F^{DM}$. This leads a small magnetic moment to appear in a material that would otherwise have zero macroscopic magnetization.

\subsection{ The linear magnetoelectric effect and weak-ferromagnetism}
Two interesting questions concerning how the electrical polarization couples to the magnetization  have arisen in recent discussions of multiferroics. The first concerns the implications of a ferroelectric distortion inducing weak ferromagnetism~\cite{fennie08} while the second concerns ferroelectric distortion inducing linear magnetoelectric effect. 
What becomes apparent after a simple Landau exercise is that two different systems with identical magnetic point groups in a ferroelectric-magnetic phase can display qualitatively different coupling physics between the {\it equilibrium} magnetization and {\it equilibrium} polarization (as opposed to small changes in the magnetization and small changes to the polarization, i.e., the response to small fields, which must be the same since they have identical magnetic point groups).
 We will see how in one case an  increase in the amplitude of the equilibrium polarization will lead to an increase in the magnitude of the equilibrium magnetization, but not in the linear magnetoelectric effect, while in a second case an increase in the equilibrium polarization leads to the exact opposite situation.
All of this  becomes apparent  by considering the relevant paraelectric reference structure and asking the question  whether the PE structure allows a linear magnetoelectric effect or the effect
 of weak-ferromagnetism, which in a paraelectric structure  are mutually exclusive.

\subsubsection{Approach: the paraelectric reference structure}

In the present context, considering a paraelectric reference structure is extremely illuminating. This 
is because in a paraelectric phase, the linear magnetoelectric effect and weak-ferromagnetism 
are mutually exclusive by symmetry~\cite{turov94, borovik-romanov06}. To see this it is clearest to revert to Landau's 
original formulation of the linear ME effect $\mathcal{F}({\it E,H}) = -\alpha_{ij} E_i H_j$ where  
{\it E} and {\it H} are the applied fields. It is obvious that if the magnetic point group contains 
either space inversion symmetry, $\mathcal{I}$, or time-reversal symmetry, $R$, then the 
linear magnetoelectric effect is zero. In a paraelectric phase, i.e., a phase containing a symmetry element
that takes $P\rightarrow -P$ and thereby enforces $P=0$,  the only allowed symmetry element 
that is compatible with the presence of a linear magnetoelectric effect is the composite operation 
of time-reversal $R$ followed by space inversion, or $R\mathcal{I}$. If the magnetic point 
group contains $R\mathcal{I}$, it displays a linear magnetoelectric effect, yet a net magnetization 
is forbidden as the system must remain invariant under $R\mathcal{I}$: $R\mathcal{I}$\MM$\rightarrow$ -{\bf M}, therefore {\bf M} =0.

In a ferroelectric phase, where there is no required symmetry that enforces $P\rightarrow -P$, both 
the linear magnetoelectric effect and weak-ferromagnetism are possible. A Landau argument, 
however, tells us nothing about the origin of these terms and hence nothing on the effects of tuning. Yet, if one considers a paraelectric reference structure for the particular ferroelectric phase of interest, a lot can be learned. For example, if in the PE reference structure the linear ME effect is allowed, then the  introduction of the ferroelectric distortion does not qualitatively change this, and therefore one 
doesn't expect the linear ME effect to scale linearly with the amplitude of the inversion-symmetry
 breaking distortion. On the other hand, in this PE reference structure, weak-ferromagnetism
 is forbidden. The ferroelectric distortion then qualitatively changes the situation, and one 
 does expect that as the size of the inversion-symmetry breaking distortion is increased, the 
 size of the weak-ferromagnetism increases. The exact opposite conclusions are reached if
 the PE reference structure of interest displays weak-ferromagnetism,
 rather than the linear ME effect.

\subsubsection{The result}
 
 We wish to search for an antiferromagetic-paralectric reference structure having a group-subgroup relation to the FM-FE phase of interest. There are two cases  that can arise.
 
 {\bf (1) }The AFM-PE structure allows a linear magnetoelectric term, $\xi$,
\begin{equation}
\mathcal{F}_A(P,M) = {1\over 2} a_p P^2 + {1\over 2} a_m M^2 + {1\over 4} b_p P^4   - \xi L_0 P M\\   
\label{PE_AFM}
\end{equation}
Since we are considering only the situation where a magnetization arises due to the coupling to some other 
order parameter, \textit{i.e.,} $a_m>0$, we only need to consider $\mathcal{O}$($M^2$).

We now consider the value of the magnetization for a fixed value of the electrical polarization and find
 $\partial\mathcal{F}_A/\partial M  =  a_m M - \xi L_0 P=0$ 
 $\Rightarrow M  =  \xi L_0 P/a_m$, and then consider small changes of $P$ due to an applied electric field, that is,
 P $\rightarrow$ $P_0$ + $\Delta P$ $$\Rightarrow M = M_0 + \Delta M$$ with
 \begin{equation}
M_0 = \left(\frac{\xi L_0}{a_m}\right)\,P_0  \,\,\,\, {\rm and}\,\,\,\,   \Delta M = \left(\frac{\xi  L_0}{a_m}\right)\,\Delta P .
\end{equation}
We see therefore that the equilibrium magnetization, $M_0$ is proportional the equilibrium value of the electrical polarization $$M_0 \propto P_0.$$ The polarization vanishes in the paraelectric phase (if the paraelectric phase exists) and is directly proportional to the ``size'' of the ferroelectric distortion. Notice that if the direction of the equilibrium electrical polarization, $P_0$, switches to its symmetry equivalent direction the equilibrium magnetization, $M_0$, switches 180$^{\circ}$ as well, and {\it this information is not accessible from a Landau theory where the expansion is about the ground state ferroelectric phase.}  The second thing to notice is that a linear magnetoelectric effect, $\Delta M\sim\Delta P$, exists whether or not the polarization is nonzero.

{\bf (2)} In the second case of interest the AFM-PE structure allows a weak-ferromagnetic term, $d$,
\begin{equation}
\label{PE_AFM_B}
\mathcal{F}_B(P,M) = {1\over 2} a_p P^2 + {1\over 2} a_m M^2 + {1\over 4} b_p P^4   + d L_0  M - \xi' L_0 P^2 M
\end{equation}
We again consider the value of the magnetization for a fixed value of the electrical polarization and find
$\partial\mathcal{F}_B/\partial M  =  a_m M + d L_0 -\xi' L_0 P^2=0$
$\Rightarrow M  =  -d L_0/a_m -\xi' L_0 P^2/a_m$,
and then considering the small changes of $P$ due to an applied electric field (P $\rightarrow$ $P_0$ + $\Delta P$) we get
$$\Rightarrow M = M_0 + \Delta M$$
with
\begin{eqnarray}
M_0 &=&  -\frac{d L_0}{a_m} + \left(\frac{\xi' L_0}{a_m}\right) P_0^2  \,\,\,\,{\rm and}\\
\Delta M &= &  \left(\frac{2\xi' L_0}{a_m}P_0\right) \Delta P \nonumber
\end{eqnarray}
We now see that the equilibrium magnetization, $M_0$ is non-zero even in the case when the 
equilibrium value of the electrical polarization vanishes (that is, in the paraelectric phase, if the paraelectric phase exists). 
 Notice that if the direction of the electrical polarization switches to its symmetry
equivalent direction the magnetization does NOT switch. {\it Again, this information is not accessible from a Landau theory where
the expansion is about the ground state ferroelectric phase.}  The second thing to notice is that a linear magnetoelectric 
effect becomes nonzero when the polarization switches on and increases in magnitude with P$_0$.

\bibliographystyle{model1-num-names}

\end{document}